\documentstyle[12pt]{article} \textwidth 14.5cm \textheight 
20cm
\begin{document} 
\newcommand{\add}{\addtocounter{eqncnt}{1}}
\newcounter{eqncnt}[section]
\newcommand{\al}{\mbox{$\alpha$}}
\newcommand{\med}[1]{g_{#1}} 
\newcommand{\meu}[1]{g^{#1}}
\newcommand{\be}{\begin{equation}\add} 
\newcommand{\ee}{\end{equation}}
\newcommand{\bea}{\begin{eqnarray}}
\renewcommand{\theequation}{\thesection.\theeqncnt} 
\newcommand{\eea}{\end{eqnarray}\add}
\newcommand{\und}{\underline}
\newcommand{\D}{\displaystyle}
\newcommand{\hs}{\hspace*{2in}}
\newcommand{\hqq}{\hfill\qquad}
\newcommand{\noin}{\noindent}
\setlength{\baselineskip}{20pt}

\title{A COMPLETE CLASSIFICATION OF SPHERICALLY SYMMETRIC PERFECT FLUID SIMILARITY SOLUTIONS}
 
\author{B. J. Carr\\
Astronomy Unit, Queen Mary \& Westfield College,\\ Mile End Road, London E1 4NS, England \\
A. A. Coley\\
Department of Mathematics \& Statistics, Dalhousie University,\\ Halifax, Nova Scotia, Canada B3H 3J5}
\maketitle
\begin{abstract}

We classify all spherically symmetric perfect fluid 
solutions of Einstein's equations with equation of state $p=\alpha \mu$ which are self-similar in the sense that all dimensionless variables depend only upon $z\equiv r/t$.
For a given value of $\alpha$, such solutions are described by two parameters and they can be classified in terms of their
behaviour at large and small distances from the origin; this usually corresponds to large and small values of $z$ but (due to a coordinate anomaly) it may
also correspond to finite $z$. We base our analysis on the demonstration (given elsewhere) that all similarity solutions must be asymptotic to solutions which depend on either powers of $z$ or powers of $lnz$. We show that there are only three similarity solutions which have an {\it exact} power-law dependence on $z$: the flat Friedmann solution, a static solution and a Kantowski-Sachs solution (although the latter is probably only physical for $\alpha <0$).  For $\alpha >1/5$, there are also two families of solutions which are asymptotically (but not exactly) Minkowski: the first is asymptotically Minkowski as $z \rightarrow \infty$ and is described by one parameter;  the second is asymptotically Minkowski at a finite value of $z$ and is described by two parameters.

As a simple application of this classification, we then present a complete analysis of the dust ($\alpha=0$) solutions,
since these can be written down explicitly and elucidate the link between the $z>0$ and $z<0$ solutions. 
As in the $\alpha \neq 0$ case, the most general dust solution is described by two parameters. The first one (E)
corresponds to the asymptotic energy at large $|z|$, while the second one (D) specifies the value of $z$ at the 
singularity which characterizes such models. The 1-parameter family with $z>0$ and $D=0$ are 
inhomogeneous cosmological models which expand from a Big Bang singularity at $t=0$ and are asymptotically
Friedmann at large $z$; models with $E>0$ are underdense and expand faster than Friedmann, while those with 
$E<0$ recollapse to black holes and contain another singularity. The $D=0$
solutions with $z<0$ are just the time reverse of the $z>0$ ones. The 
2-parameter solutions with $D>0$ again represent inhomogeneous models but they necessarily
involve both $z<0$ and $z>0$ regimes, the Big Bang singularity is at $|z|=1/D$ and (while there is no exact 
static solution in the dust case) they are asymptotically ``quasi-static'' at large $|z|$. The  
solutions with $E\ge 0$ expand or contract monotonically, whereas the ones
with  $E<0$ recollapse to or expand from a second singularity. Depending on the values of $E$ and $D$, this may be either a black
hole singularity or a naked singularity. The $D<0$ models either collapse to a shell-crossing singularity 
and become unphysical or expand from such a state. 

We then discuss
solutions with pressure. These share many
of the characteristics of the dust solutions but they also exhibit new features.
At large distances from the origin, we show that there is a 1-parameter family of
solutions which are asymptotically Friedmann, a 1-parameter family of solutions which 
are asymptotically Kantowski-Sachs and a 2-parameter family which are asymptotically quasi-static.
All these solutions can be described by parameters $E$ and $D$, analogous to those which
arose in the dust case. The asymptotically Minkowski solutions, which arise for $\alpha >1/5$, are discussed in detail elsewhere.
The possible behaviours at
small distances from the origin depend upon whether or not the solutions pass through a sonic point.
If the solutions remain supersonic everywhere, the origin corresponds to either a black hole singularity 
or a naked singularity at finite $z$, as in the dust case. However, if the
solutions pass into the subsonic region, they reach $z=0$ and their form is restricted by the requirement 
that they be regular at the sonic point.
There is again a 1-parameter family of asymptotic Friedmann solutions: this includes a continuum of underdense 
solutions and discrete bands of overdense ones; the latter are all nearly static close to the sonic point and exhibit 
oscillations. There is also a 
1-parameter family of asymptotically Kantowski-Sachs solutions. However, there are no asymptotically static 
solutions besides the exact  static solution itself.

\end{abstract}
\setcounter{equation}{0}
\section{Introduction}

Self-similar models have proved very useful in General Relativity because the similarity assumption reduces the complexity of the partial differential equations. Even greater simplification is achieved if one has spherical symmetry (Cahill \& Taub 1971) since the governing equations then reduce to 
comparatively simple ordinary differential equations. In this case, the solutions can be put into a form in which every dimensionless variable is a function of some dimensionless combination of the cosmic time coordinate $t$ and the comoving radial coordinate $r$. In the simplest situation, a similarity solution is invariant under the transformation $r \rightarrow at, t \rightarrow at$ for any constant $a$. Geometrically this corresponds to the existence of a homothetic Killing vector and is sometimes termed similarity of the ``first'' kind. We confine attention to such solutions in this paper. We shall also focus on the case in which the
source of the gravitational field is a perfect fluid with an equation of state of the form $p = \alpha \mu$. Indeed, Cahill \& Taub (1971) showed that this is the only barotropic equation of state compatible with the similarity assumption. We will assume $|\alpha| \le 1$, as required by causality, and usually require 
$\alpha$ to be positive. Note that ``geometric'' self-similarity (a property of the metric) and ``physical'' self-similarity (a property of the fluid) coincide for a perfect fluid but this need not be the case in general (Coley \& Tupper 1989). 

What makes such solutions of more than mathematical interest is the fact that they are often relevant to the real world. For example, an explosion in a homogeneous background produces
fluctuations which may be very complicated initially but which
tend to be described more and more closely by a spherically
symmetric similarity solution as time evolves (Sedov 1967). This
applies even if the explosion occurs in an expanding cosmological 
background (Schwartz et al. 1975, Ikeuchi et al. 1983). 
The evolution of cosmic voids may also be described by a similarity solution at
late times (Bertschinger 1985). The same idea applies in a wide range of contexts in fluid dynamics since, in this case, self-similar asymptotics
can be obtained from dimensional considerations. Indeed it was in this context that the concept of self-similarity of the first kind
was first introduced (Barenblatt \& Zeldovich 1972). Recently it has become clear that spherically symmetric self-similar solutions
also play a crucial role in the context of the ``critical" phenomena discovered in
gravitational collapse calculations (Choptuik 1993, Evans and Coleman 1994, Gundlach 1995, Koike et al. 1995, Maison 1996).

In the cosmological context these considerations led Carr (1995) to propose the ``similarity hypothesis''. This says that - under certain circumstances (eg. non-zero pressure and high non-linearity) - cosmological solutions may naturally evolve to a self-similar form even if they start out more complicated. There is evidence for this in both the spherically symmetric context (Carr \& Coley 1998a) and the spatially homogenous context (Wainwright \& Ellis 1997). The possibility that self-similar models may be singled out in this way from more general spherically symmetric solutions means that it is essential to understand the full family of such solutions. Unless the pressure is zero, similarity solutions generally have a shock or sonic point and the nature of the solution at this point plays a crucial role. The original study of Cahill and Taub (1971) focussed on solutions with shocks, whereas subsequent authors (eg. Bogoyavlenski 1977) have focussed on solutions with sound-wav!
es. In this paper we will only consider the latter and we will focus exclusively on
solutions which are ``regular" at the sonic point in the sense that they have a finite pressure gradient and can be continued beyond there. Even some of these solutions
will turn out to be unphysical.

Due to the existence of several preferred geometric structures in self-similar
spherically symmetric models, a number of
natural approaches (i.e. coordinate systems) may be used in studying them (Bogoyavlensky
1985). The three most common ones are the ``comoving'', ``homothetic'' and
``Schwarzschild'' approaches. In the comoving approach, pioneered by Cahill \& Taub (1973) and employed by Carr and Henriksen and coworkers, the coordinates
are adapted to the fluid 4-velocity vector. This probably affords
the best physical insights and is the most convenient one with which to study the solutions explicitly. In the homothetic approach, used by  
Bogoyavlensky and coworkers, and adopted more recently by Brady (1994) and
Goliath et al. (1998a, 1998b), the coordinates are adapted to the homothetic vector. In this case, the governing
equations reduce to those of an autonomous system and so dynamical systems
theory can be exploited to study the equations mathematically. 
The ``Schwarzschild'' approach, adopted by Ori \& Piran (1990) and Maison (1996), is useful if one
wishes to match a self-similar interior region to a non-self-similar asymptotically flat exterior region. This is because one can analyse null geodesics most simply in these coordinates, enabling the
causal structure of spacetime to be studied. The relationship between these different approaches is discussed
in more detail in Appendix A. All of them are complementary and which is most suitable depends on what type of problem one is studying. In this paper it is most convenient to use the comoving approach. 

The intention of this paper is to provide a complete classification of perfect fluid spherically symmetric self-similar solutions. We achieve this by analysing all possible 
behaviours at large and small distances from the origin. In the simplest situation this
just corresponds to large and small values of the similarity variable $z$ but the analysis is complicated by the fact that  (due to a coordinate anomaly) a finite value of $z$ may sometimes correspond to zero or infinite distance from the origin. For this reason some of the similarity solutions we discuss were missed in previous treatments (and indeed in an earlier version of this paper). A more rigorous demonstration that our
classification is complete is given elsewhere (Carr \& Coley 1998b) and consists of two parts: (1) an
analysis of all solutions whose asymptotic behaviour is associated with large or small values of $|z|$ 
and a demonstration that these always have a power-law dependence on
$z$; (2) an analysis of solutions whose asymptotic behaviour is associated with a {\it finite} value of $z$ and a demonstration that these have a power-law dependence on $lnz$. We will use this ``power-law" property as the starting point of the present analysis. This simplifies the discussion considerably and
allows us to focus on  the nature and physical significance of the solutions.
Some of them have been found before [see Coley (1997), Carr (1998) and Carr \& Coley (1998a) for recent reviews], so this will require 
some overview of previous work. However, this is the first time they have all been
brought together, with the connection between them being made explicit. This work
complements the analysis of Goliath et al. (1998a, 1998b), which also
delineates the different types of solutions but without making their physical significance clear. The precise relationship between our two approaches is discussed
in a separate paper (Carr et al. 1998).

We will show that perfect fluid self-similar spherically symmetric solutions have four possible behaviours at large distances from the origin. They are either asymptotically Friedmann, asymptotically ``quasi-static", asymptotically
Kantowski-Sachs or asymptotically Minkowski, with the last family being subdivided into two (one of which is associated with a finite value of $z$). The possible behaviours at
small distances depend upon whether or not the solutions pass through a sonic point and reach $z=0$. If the solutions remain supersonic everywhere, the origin is at finite $z$ and corresponds to either a black hole singularity or a naked singularity; in either case, the small-scale behaviour 
is uniquely determined by the large-scale behaviour. If the solutions pass through a sonic point, they may be discontinuous there and the situation is more complicated. However, in this paper we confine attention to solutions 
which are regular at the sonic point and physically realistic throughout the subsonic regime; this excludes solutions which have shocks or enter a negative mass regime. All such solutions reach $z=0$ and are either asymptotically Friedmann, exactly static or asymptotically Kantowski-Sachs for small $z$. Which of these situations 
applies depends on the value of $z$ at the sonic point and this itself is 
determined by the large-scale behaviour. In all cases, one can therefore classify solutions by their form at large distances from the origin. By way of introduction,
we now briefly describe these forms.
   
The first class of solutions is a 1-parameter family asymptotic to the flat Friedmann 
solution at large values of $|z|$. Attention originally focussed on models containing black holes because there was
interest in whether black holes could grow at the same rate as the particle horizon. Carr \& Hawking (1974) showed that such solutions
exist for radiation ($\alpha=1/3$) and dust ($\alpha=0$) 
but only if the universe is asymptotically rather than exactly 
Friedmann (i.e. there is no solution in which a black hole interior is attached to an exact 
Friedmann exterior via a sound-wave) and this has the important implication that black holes formed through purely local processes
cannot grow as fast as the Universe. [In fact, {\it all} subsonic solutions which can be attached to an exact
Friedmann model via a sound-wave are unphysical in the sense that,
as one goes inward from the sonic point, they either 
enter a negative mass regime or reach another sonic point at which
the pressure gradient diverges (Bicknell \& Henriksen 1978a).]
Carr (1976) and Bicknell \& Henriksen 1978a) then extended this result to a general $0<\alpha<1$ fluid, while Lin et al. (1976) and Bicknell \& Henriksen (1978b) considered the case of a stiff fluid ($\alpha=1$). It seems likely that all black
hole solutions are supersonic everywhere, though this has still not been rigorously proved.

There are also perturbations of flat
Friedmann models which represent density fluctuations 
growing at the same rate as the Universe (Carr \& Yahil 1990). These  solutions are asymptotically Friedmann at both large and small values of $|z|$ and regular at the sonic point. Such transonic solutions can be either
underdense or overdense relative to
the exact Friedmann model. The underdense solutions may be 
relevant to the existence of large-scale cosmic voids 
(Carr \& Whinnett 1998), while the overdense ones may be relevant to the 
formation of highly overdense blobs at a cosmic phase transition.
While there is a continuum of underdense solutions which are regular 
at the sonic point, regular overdense solutions only occur in successive and very narrow bands. These solutions have the characteristic that they are all
approximately static near the sonic point, although they depart from the static solution as they approach the origin.

The second class of models is associated with the Kantowski-Sachs solution. This is a type of homogeneous model
first studied by Kantowski and Sachs (1966) for the $\alpha =0$ case
and by Collins (1977) for arbitrary $\alpha$. For each $\alpha$ there is a unique self-similar Kantowski-Sachs solution
and there also exists a $1$-parameter family
of solutions asymptotic to this at both large and small values of $|z|$ (Carr \& Koutras 1992). Solutions with $-1/3 < \alpha <1$ are probably unphysical because they are tachyonic and the mass is negative.
Solutions with $-1< \alpha < -1/3$ may be more physical since they avoid these features. Although such equations of state violate the strong energy 
condition, they could could well arise in the early Universe due to inflation or particle production effects. Such models may be related to the growth of $p>0$ bubbles formed at a phase transition in a $p<0$ cosmological background (Wesson 1986, 1989, Ponce de Leon 1988, 1990). Note that this is the {\it only} context in which we will consider negative values of $\alpha$.

The third class of models are asympototic to either a self-similar static model or what we term ``quasi-static" models at large values of $|z|$. There is just one exactly static self-similar solution for each (positive) value of $\alpha$ (Misner \& Zapolsky 1964) and 
there is a 1-parameter family of solutions asymptotic to this. However, we will show that there is a 2-parameter family of solutions which 
are asymptotically ``quasi-static'' in the 
sense that they have an isothermal density profile at large values of $|z|$. Since the asymptotically Friedmann and asymptotically Kantowski-Sachs solutions are described by only one parameter, such solutions play a crucial role in
understanding the full family of similarity solutions. Some of these solutions remain supersonic everywhere and reach a singularity at finite $z$. However, unlike the Friedmann case, the singularity may be naked. The solutions which reach a sonic point may be attached either to the exactly static solution or the asymptotically Friedmann solutions. Some asymptotically quasi-static solutions have been studied before (Ori \& Piran 1990, Foglizzo \& Henriksen 1993). In particular, they are known to be associated with the formation of naked singularities in spherically symmetric collapse (Henriksen \& Patel 1991, Lake 1992, Joshi \& Dwivedi 1993).
However, the precise relationship of these solutions to the more general quasi-static family has not been discussed before.

The fourth class of solutions, which only exist for $\alpha >1/5$, are asymptotically Minkowski and have not been previously analysed at all. They were originally found numerically by Goliath et al. (1998b) and this led us to ``predict" them analytically. There 
are actually two such families and they are described in more detail 
elsewhere (Carr et al. 1998). Members of the first family are described by one parameter and are asymptotically Minkowski as $|z|\rightarrow \infty$; members of the second family are described by two parameters and are
asymptotically Minkowski as $z$ tends to some finite value (though this corresponds to an infinite physical distance). The existence of these solutions has been obscured until now because one needs to perform
a complicated coordinate transformation in order to put the metric into an explicitly 
Minkowski form. As in the asymptotically Friedmann and asymptotically quasi-static cases, these solutions may either be supersonic everywhere (in which case they contain a black hole or naked singularity) or they may be attached
to $z=0$ via a sonic point (in which case they are asymptotically Friedmann at small $z$). The latter solutions are particularly interesting because they are both associated with the occurence of critical phenomena, as discussed in more detail by Carr et al. (1998) and Carr, Henriksen \& Levy (1998).

The plan of this paper is as follows. In Section 2 we will introduce the relevant equations and discuss the crucial role of the sonic point. In Section 3 we will analyse the possible behaviours at large and small distances from the origin, 
emphasizing the key role played by the power-law and log-power-law solutions. In Section 4 we will analyse the dust $(p=0)$ solutions since the complete family of such solutions can be derived analytically. Despite the simplifications entailed in dropping pressure,
we will find that many of their features carry over to 
the $\alpha \neq 0$ case. In particular, the dust solutions illuminate the connection between
models with positive and negative $z$. In Section 5 we will consider the solutions with pressure. We will show that some of the features of these solutions can be understood by combining the insights gained from the dust solutions in the supersonic regime with those gained from the asymptotically Friedmann solutions in the subsonic regime. However, other kinds of solution have no analogue in the dust case. We make some final remarks in Section 6.
    
\setcounter{equation}{0}
\section{Spherically Symmetric Similarity Solutions}

In the spherically symmetric situation one can introduce a time coordinate
$t$ such that surfaces of constant $t$ are orthogonal to fluid flow lines
and comoving coordinates ($r,\theta ,\phi$) which are constant along each
flow line. The metric can then be written in the form
\be
\label{lelement}
        ds^{2}=e^{2\nu}\,dt^{2}-e^{2\lambda}\,dr^{2}-R^{2}\,d\Omega^{2},\;\;\;
        d\Omega ^2 \equiv d\theta^{2}+\sin^{2}\theta \,d\phi^{2} 
\ee
where $\nu$, $\lambda$ and $R$ are functions of $r$ and $t$. For a perfect 
fluid the Einstein equations are
\be
        G^{\mu \nu}=8\pi[(\mu+p)U^{\mu}U^{\nu}-p\,g^{\mu \nu}]
\ee
where $\mu(r,t)$ is the energy density, $p(r,t)$ the pressure, 
$U^{\mu}=(e^{-\nu},0,0,0)$ is the comoving fluid 4-velocity, and we choose
units in which $c=G=1$. The equations have a first integral
\be
\label{firstint}
   m(r,t)=\mbox{$\frac{1}{2}$} R\left[ 1+e^{-2\nu}
   \left(\frac{\partial R}{\partial t}
   \right)^{2} - e^{-2\lambda}\left(\frac{\partial R}
   {\partial r}\right)^{2}\right]
\ee
and this can be interpreted as the mass within comoving radius $r$ at 
time $t$: 
\be
\label{massfunct}
   m(r,t)=4\pi\int_{0}^{r}\mu R^{2}\frac{\partial R}{\partial r'}\,dr'.
\ee
Unless $p=0$, this quantity decreases with increasing $t$ because of the
work done by the pressure. One can also express it as
\be
\label{massfunct2}
   m(r,t)=4\pi\int_{0}^{t} p R^{2}\frac{\partial R}{\partial t'}\,dt'
\ee
and this is the more appropriate expression when there is no spatial origin (as in the Kantowski-Sachs solution).
Eqn (2.3) can be written as an equation for the energy per
unit mass of the shell with comoving coordinate $r$:
\be
\label{efunc}
   E\equiv\frac{1}{2}(\Gamma^{2}-1)
   =\frac{1}{2}U^{2}-\frac{m}{R},
    \;\;\;\;U\equiv e^{-\nu}\left(\frac{\partial R}{\partial t}\right),
   \;\;\;\;\Gamma\equiv e^{-\lambda}\left(\frac{\partial R}{\partial r}\right).
\ee
This can be interpreted as the sum of
the kinetic and potential energies per unit mass. Only in the $p=0$ case are $E$ and $\Gamma$ conserved along fluid flow lines. 

By a spherically symmetric similarity solution we shall mean  one in 
which the spacetime admits a homothetic Killing vector \mbox{\boldmath$\xi$} that satisfies
\be
        \xi_{\mu ;\nu}+\xi_{\nu ;\mu} =2g_{\mu \nu}.
\ee 
This means that the solution is unchanged by a transformation of the form
$t\rightarrow at$, $r\rightarrow ar$ for any constant $a$. 
Solutions of this sort were first
investigated by Cahill \& Taub (1971), who showed that by a suitable
coordinate transformation they can be put into a form in which all
dimensionless quantities such as $\nu$, $\lambda$, $E$ and
\be
   S\equiv\frac{R}{r},\;\;\;\;M\equiv\frac{m}{R},\;\;\;\;
   P\equiv pR^{2},\;\;\;\;W\equiv\mu R^{2}
\ee
are functions only of the dimensionless variable $z\equiv r/t$. Then we have
\be
   \frac{\partial\;}{\partial t}=-\frac{z^{2}}{r}\frac{d\;}{dz},\;\;
   \;\; \frac{\partial\;}{\partial r}=\frac{z}{r}\frac{d\;}{dz},
\ee
so the field equations reduce to a set of ordinary differential
equations in $z$. Another important quantity is the function
\be
\label{velocity}
        V(z)=e^{\lambda -\nu}z,
\ee
which represents the velocity of the surfaces of 
constant $z$ relative to the fluid. These surfaces have the equation $r=z t$ and therefore
represent a family of spheres moving through the fluid. The spheres contract relative to the fluid for $z<0$ and expand for $z>0$. This is to be distinguished from the velocity of the spheres of constant R relative to the fluid:
\be
V_R = - \frac{U}{\Gamma} = - e^{\lambda -\nu} \left(\frac{\partial R/\partial t}{\partial R/\partial r}\right).
\ee 
This is positive if the fluid is collapsing and negative if it is expanding.
Special significance is attached to values of $z$ for which
$|V|=1$, $V_R=0$ and $|V_R|=1$. The first corresponds to a Cauchy horizon (either a black hole event horizon
or a cosmological particle horizon), the second to a stagnation point, and the third
to a black hole or cosmological apparent horizon. We show shortly that the existence of an apparent horizon is also equivalent to the condition $M=1/2$.

The only barotropic equation of state compatible with the similarity ansatz is
one of the form $p=\alpha \mu$ ($-1\leq\alpha\leq 1$). It is convenient to 
introduce a dimensionless function $x(z)$ defined by
\be
\label{xdef}
        x(z)\equiv (4\pi \mu r^{2})^{-\alpha/(1+\alpha)}.
\ee
[Note that the factor of $4\pi$ is omitted in the definition of $x$ given
by Carr \& Yahil (1990) but it is required for consistency with eqn (2.3).] The conservation equations $T^{\mu \nu}_{\;\;\;\; ;\nu}=0$ can then be 
integrated to give
\be
\label{metrictt}
        e^{\nu}=\beta x z^{2\alpha/(1+\alpha)}
\ee
\be
\label{metricrr}
        e^{-\lambda}=\gamma x^{-1/\alpha} S^{2}
\ee
where $\beta$ and $\gamma$ are integration constants. The remaining field 
equations reduce to a set of ordinary differential equations in $x$ and 
$S$:
\be
   \ddot{S}+\dot{S}+\left(\frac{2}{1+\alpha}\frac{\dot{S}}{S}
   -\frac{1}{\alpha}\frac{\dot{x}}{x}\right)
   [S+(1+\alpha)\dot{S}]=0,
\ee
\be
   \left(\frac{2\alpha\gamma^{2}}{1+\alpha}\right)S^{4}
   +\frac{2}{\beta^{2}}\frac{\dot{S}}{S}\,x^{(2-2\alpha)/\alpha}
   z^{(2-2\alpha)/(1+\alpha)} - \gamma^{2}S^{4}\,\frac{\dot{x}}{x}
   \left(\frac{V^{2}}{\alpha}-1\right) = (1+\alpha)x^{(1-\alpha)/\alpha},
\ee
\be
   M=S^{2}x^{-(1+\alpha)/\alpha}\left[1+(1+\alpha)\frac{\dot{S}}{S}
   \right],
\ee
\be
   M=\frac{1}{2}+\frac{1}{2\beta^{2}}x^{-2}z^{2(1-\alpha)/(1+\alpha)}
   \dot{S}^{2}-\frac{1}{2}\gamma^{2}x^{-(2/\alpha)}S^{6}
   \left(1+\frac{\dot{S}}{S}\right)^{2},
\ee
where the velocity function is given by
\be
   V=(\beta\gamma)^{-1}x^{(1-\alpha)/\alpha}S^{-2}
   z^{(1-\alpha)/(1+\alpha)}
\ee
and an overdot denotes $zd/dz$. The other velocity function is
\be
V_R = \frac{V\dot{S}}{S+\dot{S}}\;\;,
\ee
while the energy is
\be
E = \frac{1}{2}\gamma^{2}x^{-(2/\alpha)}S^{6}
   \left(1+\frac{\dot{S}}{S}\right)^{2} - \frac{1}{2}\;\;.
\ee
Eqn (2.18) can then be written in the form
\be
M=\frac{1}{2}+ \left(E + \frac{1}{2}\right) (V_R^2 -1),
\ee 
so the condition $M=1/2$ implies that $|V_R|=1$ (corresponding to an apparent horizon). Note that $M \neq 1/2$ in the special case $E=-1/2$ because this corresponds to the Kantowski-Sachs solution and, for this, $V_R$ diverges.

As discussed by Carr \& Yahil (1990), we can best
envisage how these equations generate solutions by working in the
$3$-dimensional $(x, S, \dot{S})$ space. At any point
in this space, for a fixed value of $\alpha$, eqns (2.17) and (2.18) give the value of z; eqn (2.16) then gives the value of $\dot{x}$ unless $|V|=\sqrt{\alpha}$ and eqn (2.15) gives the value of $\ddot{S}$. Thus the equations generate a vector
field $(\dot{x}, \dot{S}, \ddot{S})$ and this specifies an integral curve at each point of the 3-dimensional space. Each curve
is parametrized by $z$ and represents one particular similarity solution.
This shows that, for a given equation of state parameter $\alpha$, there is a
$2$-parameter family of spherically symmetric similarity solutions.

In $(x, S, \dot{S})$ space the sonic condition $V= \sqrt{\alpha}$ specifies a $2$-dimensional surface because eqns (2.17) to (2.19) allow one to express $\dot{S}$ in terms of $x$ and
$S$. The same surface corresponds to the condition $V= -\sqrt{\alpha}$.
Where a curve intersects this surface, eqn (2.16) does not uniquely
determine $\dot{x}$, so there can be a number of different solutions passing
through the same point. However, integral curves intersect
$|V| = \sqrt{\alpha}$ in a physically reasonable manner only if
\be
\left(\frac{2\alpha\gamma^{2}}{1+\alpha}\right)S^{4}
   +\frac{2}{\beta^{2}}\frac{\dot{S}}{S}\,x^{(2-2\alpha)/\alpha}
   z^{(2-2\alpha)/(1+\alpha)} = (1+\alpha)x^{(1-\alpha)/\alpha},
\ee
since otherwise the value of $\dot{x}$ and
hence the pressure, density and velocity gradient diverge there.
Since eqn (2.23) corresponds to another 2-dimensional surface in $(x, S, \dot{S})$ space, this will intersect the surface $|V|=\sqrt{\alpha}$ on a line $Q$. Only integral curves which hit the sonic surface on this line are ``regular'' in the sense that they can be extended beyond there. (All other solutions would have to contain shock-waves.) From each point on this line there will be regular integral curves with decreasing and
increasing $z$.  One can join any member of the first kind to any member
of the second kind to obtain a complete similarity solution. 

Physically reasonable solutions cannot have an arbitrary value of $\dot{x}$ at
$|V| = \sqrt{\alpha}$.  If we require $\ddot{x}$ to be
finite there, then the equations
permit just two values of $\dot{x}$ at each point of the line $Q$ and there will then be two corresponding values of $\dot{V}$. If
the values of $\dot{x}$ are complex, corresponding to a {\it focal} point, then the
solution will be unphysical. If they are real, at least one of the values of $\dot{V}$ must be positive.  If both values of $\dot{V}$ are positive, corresponding to a {\it nodal} point, the smaller one is associated with a $1$-parameter family of solutions, while the larger one
is associated with an isolated solution.  If one of the values of $\dot{V}$ is
negative, corresponding to a {\it saddle} point, both values are associated with isolated solutions. This behaviour has been analysed in detail by Bogoyavlenski (1977), Bicknell \& Henriksen (1978), Carr \& Yahil (1990) and Ori \& Piran (1990).

One can show that there is a $1$-parameter family
of regular solutions (i.e. a node) only on a restricted part of the line $Q$ and, in the
$V(z)$ diagram, this corresponds to two ranges of values for $z$. One range $(z_1 < z < z_2)$ lies to the left of the Friedmann sonic point $z_F$ and includes the static sonic point $z_{S}$. The other goes from some value $z_3$ to infinity and includes $z_F$.
We will argue later that any family of regular solutions which is described by just one parameter asymptotically 
probably has to hit the sonic line in these ranges. One has a saddle point for $z<z_1$ and a focal point for $z_2<z<z_3$. These features are
indicated in Figure (1a). The values of $z_1$, $z_2$ and $z_3$ can be given in 
terms of $\alpha$ but the expressions are very complicated, so we do not give them explicitly. The ranges for $\alpha =1/3$ are indicated in Figure (1b); in this case, $z_3 =z_F$ and $z_2=z_{S}$. 
On each side of a nodal sonic point, $\dot{x}$ may have either of its two possible values.  If one chooses different values for $\dot{x}$, there will be a discontinuity in the pressure gradient.  If one chooses the same value,
there may still be a discontinuity in the second derivative of $x$.  Only 
the isolated solution and a single member of the $1$-parameter family of solutions  are ``analytic" or at least $C^\infty$ in the sense that derivatives of every order are continuous. This contrasts with the case of a shock where $x$ is itself discontinuous.  

\section{Asymptotic Behaviour of Similarity Solutions}

The key step in providing a complete classification of spherically symmetric perfect fluid similarity solutions is an analysis of their possible asymptotic behaviours and we
now present this. For simplicity we will assume $z>0$ 
throughout this section but the analysis can be trivially extended to the $z<0$ case. 
We will also assume $\alpha >0$ except in the Kantowski-Sachs case. The full technicalities of the asymptotic analysis are presented elsewhere (Carr \& Coley 1998b). For present purposes it suffices to note that all similarity
solutions depend on powers of $z$ at large and small values of $|z|$ or on powers of $lnz$ at finite $z$. The last possibility arises because, due to a coordinate anomaly, a
finite value of $z$ may sometimes correspond to zero or infinite physical distance. In this section we will identify these asymptotic states explicitly. We will show that there are three {\it exact} power-law solutions: the flat Friedmann solution, a Kantowski-Sachs solution and a static solution. We will also show that, for $\alpha>1/5$, there are asymptotically (but not exact) Minkowski solutions which asymptote either to infinite
$z$ or finite $z$. Finally there are solutions
whose origin corresponds to a singularity at finite $z$. The validity of these results is confirmed by the dynamical systems analyses of Bogoyavlensky (1985) and Goliath et al. (1998a, 1998b). In particular, the
existence of the monotone and Dulac functions found in these analyses forbids the existence
of periodic orbits and limit cycles and thereby excludes other possible asymptotic
behaviours.

\subsection{Power-Law Similarity Solutions}

In order to find the asymptotically ``power-law" solutions explicitly, we look for solutions to the field equations of the form
\be
   x=x_{o} z^{a},\;\;\;S=S_{o} z^{b}
\ee
where $x_{o}$, $S_{o}$, a and b are constants. Note that $\dot{S}/S=b$ and $\dot{x}/x=a$. Eqn (2.15) is satisfied if
\be
a = \frac{b\alpha [3(b+1) + \alpha (3b+1)]}{(1+\alpha)[1+(1+\alpha)b]}.
\ee
The factor $[1+(1+\alpha)b]$ cannot be zero since
this would be inconsistent with eqns (2.17) and (2.18). Eqn (2.16) can then be written in the form
\be
Az^p + Bz^q + C = 0
\ee
where
\bea
A  \equiv  \frac{b[(\alpha-1) + (1+\alpha) (2 \alpha -1)b]}{\beta^2(1+\alpha)[1+(1+\alpha)b]} \, x_0^{2(1-\alpha)/\alpha} S_0^{-4},\;\;\;
B \equiv -(1+\alpha) x_0^{(1-\alpha)/\alpha}  S_0^{-4}, \nonumber\\  
C  \equiv  \frac{\alpha \gamma^2 (b+1) [2 + 3b(1+\alpha)]}{(1+\alpha)[1+(1+\alpha) b]} \;\;\;
\eea
and the exponents are
\be
 p \equiv 2a \left( \frac{1-\alpha}{\alpha}\right) - 4b + 2\left( \frac{1-\alpha}{1+\alpha}\right),\;\;\;     \nonumber
q \equiv a\left( \frac{1- \alpha}{\alpha}\right) -4b.
\ee
Since B cannot be zero, there are then three ways in which eqn (3.3) can be satisfied to leading order as $z \rightarrow 0$ or $z \rightarrow \infty$ and we discuss these in turn.

\vskip .3in
$\bullet$ $p=q, A+B=0$. In this case, the condition $p=q$ implies
\be
   a=-\frac{2\alpha}{1+\alpha}
\ee
and the condition $A+B=0$ implies
\be
   x_{o}^{(1-\alpha)/\alpha}=\frac{1}{2}\beta^{2}\left[\frac{(1+\alpha)^{2}}
   {b(1+\alpha)+1}\right] .   
\ee
Eqn (3.2) then requires
\be
b=-1\;\;\; or\;\;\; -\frac{2}{3(1+\alpha)}
\ee
and both values lead to $C=0$ from eqn (3.4). Since eqn (3.3) is satisfied exactly, there are no approximate solutions with $C\neq 0$.

The choice $b=-2/[3(1+\alpha)]$ corresponds to the flat  Friedmann model. In this case, eqns (2.17) and (2.18) are satisfied if 
\be
   x_{o}^{(\alpha -1)/\alpha}=\frac{2}{3\beta^{2}(1+\alpha)^{2}},\;\;\;\;
   \gamma^{2}S_{o}^{6}x_{o}^{-2/\alpha}=\frac{9(1+\alpha)^{2}}{(1+3\alpha)^{2}},
\ee
and one can choose $x_{o}=S_{o}=1$ providing one scales the $r$
and $t$ coordinates such that
\be
\label{intconst}
        \beta=\frac{\sqrt{2}}{\sqrt{3}(1+\alpha)},\;\;\;\;
        \gamma=\frac{3(1+\alpha)}{(1+3\alpha)}.
\ee
This gives 
\be
        x=z^{-2\alpha/(1+\alpha)},\;\;\;\;
        S=z^{-2/[3(1+\alpha)]}
\ee
and the metric becomes
\be
   ds^{2}=\beta^{2}\,dt^{2}-\gamma^{-2}z^{-4/[3(1+\alpha)]}\,dr^{2}-
   r^{2(1+3\alpha)/[3(1+\alpha)]}\,t^{4/[3(1+\alpha)]}d\Omega^{2}.
\ee
One can put
it in a more familiar form by making the coordinate transformation
\be
   {\hat{t}}=\beta t,\;\;\;\;
   {\hat{r}}=\beta^{-2/[3(1+\alpha)]}r^{(1+3\alpha)/[3(1+\alpha)]},
\ee
which gives
\be
   ds^{2}=d{\hat{t}}\;^{2}-{\hat{t}}^{4/[3(1+\alpha)]}[d{\hat{r}}^{2}+
   {\hat{r}}^{2}d\Omega^{2}].
\ee
This is just the flat  Friedmann solution with $p=\alpha \mu$. 
We also have
\be
\label{friedvel}
        \mu = \frac{1}{4\pi t^2},\;\;\;
        V=\left(\frac{1+3\alpha}{\sqrt{6}}\right)
        z^{(1+3\alpha)/[3(1+\alpha)]},\;\;\;
        M=\mbox{$\frac{1}{3}$}z^{2(1+3\alpha)/[3(1+\alpha)]}.
\ee

The choice $b=-1$ corresponds to the self-similar Kantowski-Sachs (KS) model. This is compatible with eqn (3.7) providing
\be
   \beta^{2}=-\frac{2\alpha}{(1+\alpha)^{2}}x_{o}^{(1-\alpha)/\alpha}.
\ee
>From eqns (2.17) and (2.18) we also require
\be
   S_{o}^{2}x_{o}^{-(1+\alpha)/\alpha}= \frac{2\alpha}{(1+\alpha)^{2}-4\alpha^{2}}.
\ee
Eqn (3.17) shows that we cannot take $x_{o}=S_{o}=1$ in this case but both $x_{o}$ and $S_{o}$ are determined in terms of $\alpha$ and $\beta$. The constant $\gamma$ is not constrained at all. If we take $\beta$ and $\gamma$ to have the values given by eqn (3.10) for $\alpha<0$ and $i$ times those values for $\alpha>0$, so that we have the same $r$ and $t$ scaling as in the Friedmann solution, then eqns (3.16) and (3.17) give 
\be
   x_0 = \left(\frac{1}{3 |\alpha|}\right)^{\alpha/(1 - \alpha)},\;\;\;
   S_0 ^2 = \frac{2\alpha}{(1 + 3 \alpha)(1 - \alpha)} 
   \left(\frac{1}{3 |\alpha|}\right)^{(1 + \alpha)/(1 -\alpha)}.
\ee
[Carr \& Koutras  (1993) do not incorporate the $i$ factors for $\alpha>0$ but this is a less sensible normalization since it allows the metric to be complex.] We now have
\be
S=S_0z^{-1}, \;\;\; x=x_0z^{-2\alpha/(1+\alpha)}
\ee
and the metric is
\be
   ds^2 = \beta^2 x_0^2dt^2 - \gamma^{-2}x_0^{2/\alpha}S_0^{-4}
   z^{4 \alpha/(1 + \alpha)} dr^2 - S_0^2 t^2 d\Omega ^2.
\ee
The $t$ coordinate is spacelike and the $r$ coordinate is timelike for $\alpha>0$ because of the $i$ factors in $\beta$ and $\gamma$.  For $-1/3<\alpha<0$, $t$ and $r$ have their usual interpretation but, from eqn (3.18), the circumferential coordinate is timelike since $S_{o}$ is imaginary. One can put the metric in a more familiar form by making the coordinate transformation
\be
   \hat{t}= \beta x_0 t,\;\;\; \hat{r}= \gamma^{-1}
   (\beta x_0)^{2\alpha/(1 + \alpha)} x_0\,^{1/\alpha} S^{-2}_0\,
   r\,^{(1 + 3 \alpha)/(1 + \alpha)},
\ee
which gives 
\be
  ds^2 = d\hat{t}^2 - \hat{t} ^{-4 \alpha/(1 + \alpha)} d \hat{r}^2
  - (S_0/\beta x_0)^2\hat{t}^2 d \Omega^2. 
\ee
This corresponds to a $p = \alpha \mu$ KS
solution. We also have
\bea
   \mu t^{2} =  \left( \frac{1}{3 |\alpha|}\right)^{(1+\alpha)/(\alpha-1)},\;\;
   V = -\frac{(1-\alpha)(1+3\alpha)^2}{2 \sqrt{6}\alpha} \left( \frac{1}{3|\alpha|}\right)^{-2 \alpha/(1-\alpha)} z^{(1+ 3 \alpha)/(1+\alpha)},         \nonumber\\
   M = \frac{2 \alpha^2}{(\alpha-1)(3 \alpha+1)} \equiv M_{KS}.\;\;\; . 
\eea
$V$ is negative for $0<\alpha<1$ (corresponding to tachyonic solutions), while $M$ is
negative for $-1/3<\alpha<1$ (corresponding to negative mass solutions).  Solutions with $\alpha < -1/3$ are therefore more physical in 
that $V$ and $M$ are positive;
although such solutions have negative pressure and violate the strong
energy condition, the required conditions might arise naturally in the early Universe. Note that eqn (2.4) does not apply in this case
because there is no well-defined origin; eqn (3.19) implies that $R$ is independent of $r$, so everything is on a shell. Instead the
value of $m$ must be interpreted as the mass of the 
whole Universe at time $t$, as indicated by eqn (2.5).

\vskip .3in
$\bullet$ $ q=0, B+C=0$. In this case, one can show that the only consistent solution is
\be
 a=b=0,
\ee
i.e. $x$ and $S$ are constant. [The condition $q=0$ permits another value of $b$ but
this leads to negative $C$, so the condition $B+C=0$ cannot be satisfied.] Eqn (3.24) implies that $A$ is zero and hence eqn (3.3) is satisfied
identically, so there are no approximate solutions with $A \neq 0$. The condition $B+C=0$ also requires
\be
   S_{o}^{2} = \frac{1+\alpha}{\gamma\sqrt{2\alpha}} x_o^{(1-\alpha)/2\alpha}
\ee
for $\alpha \neq 0$. This corresponds to the exact self-similar static solution, with the metric being given by 
\be
   ds^2=\beta^2 x_o^2 z^{4\alpha/(1+\alpha)}dt^2 - \gamma^{-2} x_o^{2/\alpha} S_o^{-4}dr^2 - r^2 S_o^2 d\Omega^2.
\ee
This can be put in an explicitly static form
\be
ds^2 = \hat{r}^{4\alpha/(1+\alpha)}d\hat{t}^2 - \gamma^{-2}x_o^{2/\alpha}S_o^{-6}d\hat{r}^2 - \hat{r}^2 d\Omega ^2
\ee
by introducing the variables
\be
\hat{r} = rS_o,\;\;\;\hat{t} = \left( \frac{1+\alpha}{1-\alpha} \right) \beta x_o S_o^{-2\alpha/(1+\alpha)}t^{(1- \alpha)/(1+\alpha)}.
\ee 
The other relevant functions are
\be
   \mu = x_o^{-(1+\alpha)/\alpha}r^{-2},\;\;\;
V = x_o^{-(1-\alpha)/2\alpha}z^{(1- \alpha)/(1+\alpha)},\;\;\;
    M = \frac{2\alpha}{1+6\alpha+\alpha^{2}}.
\ee
If $\beta$ and $\gamma$ have the same values as in the Friedmann solution, eqns (2.17), (2.18) and (3.25) imply that $x_o$ and $S_o$ are given by
\be
x_o = \left[\frac{9(1+3\alpha)(1+6\alpha + \alpha ^2)}{(18\alpha)^{3/2}}\right]
^{2\alpha /(1+3\alpha)},\;\;\;
S_o = \left[\frac{(1+6\alpha + \alpha ^2)^{(1-\alpha)/2}(\alpha +1/3)^{1+\alpha}}{2\alpha}\right]^{1/(1+3\alpha)}
\ee
so there is only one static solution for each equation of state. Note that eqn (2.20) implies that $V_R=0$ in this case, as expected. It should be stressed that there is no static solution in the dust case, essentially because one cannot put the $\dot{x}/ \alpha$ term in eqn (2.15) to zero. [If one puts $\alpha=0$ in 
eqns (3.29) and (3.30), one obtains $M=0$ and the spatial metric 
components all diverge; also eqn (3.25) cannot be satisfied.] The self-similar static solution was first discussed by Misner \& Zapolsky (1964) and subsequently by Henriksen \& Wesson (1978) and Carr \& Yahil (1990).

Note that there is an interesting connection between the static and KS solutions:  if one interchanges the $r$ and $t$ coordinates in metric (3.26) and also changes the equation of state parameter to
\be
   \alpha' = -\frac{\alpha}{1+2\alpha},
\ee
one obtains the KS metric (3.20). For a static solution with a normal equation of state $(1>\alpha>0)$, $\alpha'$ must lie in the range $-1/3$ to $0$, so some negative pressure KS solutions are related to positive pressure static ones. However, KS solutions with $-1<\alpha '<-1/3$ correspond to $|\alpha|>1$
and so do not give physical static solutions. Note that $\alpha=\alpha'$ only for $\alpha=0$ or $\alpha=-1$. The mass of both the static and KS solutions tends to $0$ as $\alpha\rightarrow 0$, although the solutions do not exist in the limit $\alpha =0$ itself.

\vskip .3in
$\bullet$ $p=0, A+C=0$. The condition $p=0$  implies 
\be
b = \frac{1}{2}\left(\frac{1-\alpha}{\alpha}\right)a + \frac{1}{2}\left(\frac{1-\alpha}{1+\alpha}\right)
\ee
and eqn (2.19) then requires that $V$ tend to the finite value
\be
V_* = \beta^{-1}\gamma^{-1}x_o^{(1-\alpha)/\alpha}S_0^{-2}.
\ee
The condition $A+C=0$ now implies
\be
a = \frac{V_*^2(1-\alpha)+2\alpha}{(V_*^2-1)(1+\alpha)},\;\;\;
b = \frac{(1-\alpha)(V_*^2+\alpha)}{2\alpha(1+\alpha)(V_*^2-1)},
\ee
while eqn (3.5) yields
\be
q = \frac{(1-\alpha)V_*^2}{\alpha(1-V_*^2)}.
\ee
This is only a consistent solution of eqn (3.3) for large $z$ if $V_*^2 > 1$ and for small $z$ if $V_*^2 < 1$. However, eqn (2.17) implies that the latter condition leads to negative values
of $M$ (and hence to unphysical solutions) unless $V_*^2<\alpha$.
Since $V_*^2 > 1$, eqn (3.34) implies that both $a$ and $b$ are positive, so the density goes to zero and the scale factor goes to infinity in these solutions.  It should be emphasized that, since $B \neq 0$ from eqn (3.4), this case does not lead to an {\it exact} solution. Eqn (2.17) gives
\be
M \sim z^{-[V_*^2(1-\alpha)+1+3\alpha]/(V_*^2-1)(1+\alpha)}
\ee
and this necessarily tends to zero as $z \rightarrow \infty$. On the other hand, eqn (2.18) implies
\be
M - \frac{1}{2} \sim z^{[V_*^2(1-\alpha)-\alpha(1+3\alpha)]/
(V_*^2 -1) \alpha (1+\alpha)} [b^2(V_*^2-1) - 2b - 1].
\ee
If the exponent of $z$ in this expression is positive, $M \rightarrow 0$ as $z\rightarrow \infty$ only if the term in square brackets does and this requires
$b=1/(V_*-1)$. Eqn (3.34) then gives a quadratic equation for $V_*$:
\be
 (1-\alpha) V_*^2 -2\alpha(1+\alpha)V_* - \alpha(1+3\alpha) =0
\ee
with the real positive solution
\be
V_* = \frac{\alpha(1+\alpha) + \sqrt{\alpha(\alpha^3 - \alpha^2 + 3\alpha +1)}}{1 - \alpha}.
\ee
[This assumes that $\alpha$, $V_*$ and $z$ are all positive; for $\alpha <0$ one would
need to take the negative square root in eqn (3.39).] Note that eqn (3.38) implies that the exponent of $z$ in eqn (3.37) is indeed positive (as assumed). 

The value of $V_*$ given by eqn (3.39) exceeds 1 only for $\alpha >1/5$, so these solutions do not exist in the dust case. [For $\alpha <1/5$ there are solutions as $z\ rightarrow 0$ with 
$1>V_*^2>\alpha$ but these have negative mass.] Eqns (3.33) and (3.39) impose a relationship between $x_0$ and $S_0$, so these solutions are described by just one independent parameter. Requiring that the right-hand-side of eqn (3.37) tends to $-1/2$ as $z \rightarrow \infty$ merely determines the second order terms in the expansions for $x$ and $S$.
The metric has the asymptotic form
\be
ds^2 \sim z^{2V_*^2/(V_*^2 -1)}dt^2 - z^{2/(V_*^2 -1)}dr^2 - r^2 z^{2/(V_* -1)}d\Omega^2
\ee
and this can be reduced to the Minkowski form with a suitable change of coordinates.
Note that although $MS \rightarrow 0$ as $z\rightarrow \infty$, it does so slower than 
$z^{-1}$, so that the mass itself ($m=rMS$) diverges. 

\vskip .3in
The forms of $V(z)$ for the Friedmann, static and KS 
solutions are shown in Figure (1a) for the general $\alpha$ case
and in Figure (1b) for the $\alpha =1/3$ case. The asymptotically Minkowski
solutions for $\alpha >1/5$ are not included since they are not {\it exact} (viz. Minkowski has no matter). Note also that the Minkowski solution, although static, is distinct
from the exact self-similar static solution given by eqn (3.26). A rather peculiar feature of the similarity solutions, which arose in the 
context of the KS model, is that the mass
can go negative. This may seem unphysical but - in the context of the Big Bang model - Miller (1976) has given a possible interpretation of this in terms of ``lagging'' cores. In the $\alpha =1/3$ case (and {\it only} this case), one can show that there is a 
curve in the $V(z)$ diagrams where $M=0$:
\be
   V^3 = - \sqrt{3/2}\; z^{3/2} (V^2 - 1/9).
\ee
and this is also shown in Figure (1b). One sees that the curve has
asymptotes at $V=\pm 1/3$.  The upper part (with $V > 1/3$) is relevant for
asymptotically Friedmann solutions, while the lower part (with $V < -1/3$) is 
relevant for asymptotically KS solutions. $M$ is negative in between the
two parts and this region includes KS itself (as expected). Note that eqn (3.41)
is not a {\it sufficient} condition for $M=0$; in fact, it implies that M has
two possible values, only one of which is zero.  

\subsection{Logarithmic Power-Law Similarity Solutions}

By analogy with eqn (3.1) we now look for solutions in which $z$ tends to some finite value $ z_*$ and in which
\be
x=x_0 L^a,\;\;\;S=S_0 L^b,\;\;\;L\equiv ln(z/z_*)
\ee
for constants $x_0$ and $S_0$. Clearly $z=z_*$ corresponds to an infinite distance
from the origin for $b<0$ and zero distance for $b>0$. Eqn (2.15) requires
\be
b\left[b-1+(1+\alpha)\left(\frac{2b}{1+\alpha}-\frac{a}{\alpha}\right)\right]
+ \left[ \left(\frac{3+\alpha}{1+\alpha}\right) b -\frac{a}{\alpha}\right]L = 0
\ee
and the leading term is zero only for
\be
b = \frac{1}{3} + \left( \frac{1+\alpha}{3\alpha}\right) a.
\ee
It turns out that the last term is never zero, so these are only asymptotic and not exact solutions. There are now two possible situations, according to whether $V$ tends to infinity
or some constant value $V_*$. 

\vskip .3in
$\bullet$ $V\rightarrow V_*$ as $z\rightarrow z_*$. In this case, eqn (2.16) can be written in the form
\be
aL^{-1} = \frac{2V_*^2\alpha b}{(V_*^2-\alpha)}L^{-1}+ \frac{2\alpha^2}{(1+\alpha)(V_*^2-\alpha)} -  \frac{\alpha (1+\alpha)\beta^2 V_*^2}{V_*^2-\alpha} z_*^{-2(1-\alpha)/(1+\alpha)} (x_0 L)^{a(\alpha -1 )/\alpha}. 
\ee
The only consistent solution to this equation 
has the last term tending to zero and this then implies
\be
a = \left(\frac{2V_*^2\alpha}{V_*^2-\alpha}\right)b.
\ee
However, eqn (2.19) also implies
\be
a = \left(\frac{2\alpha}{1-\alpha}\right)b,
\ee
so we require $V_*^2=1$. Eqns (3.44) and (3.47) determine $a$ and $b$ and lead to
\be
S =S_0 L^{(1-\alpha)/(1-5\alpha)},\;\;\;x = x_0 L^{2\alpha/(1-5\alpha)}.
\ee
Thus the scale factor diverges and the density goes to zero providing $\alpha >1/5$. However, there are no such solutions for $\alpha <1/5$, so they do not appear in the dust case. The condition $V_*=1$ gives a 
relationship between the constants $x_0$, $S_0$ and  $z_*$, so these solutions are
described by two independent parameters. Eqn (2.17) implies
\be
M \sim L^{(1-\alpha)/(5\alpha -1)},
\ee
so this is zero at $z=z_*$. However, eqns (3.48) and (3.49) imply that the product $MS$ tends to a constant, so the mass itself is not zero. Eqn (2.18) can be written as
\be
M = \frac{1}{2} + \frac{1}{2}\gamma^2 x^{-2/\alpha} S^6 \left\{\left(\frac{\dot{S}}{S}\right)^2(V^2-1) - \frac{2\dot{S}}{S} - 1\right\}.
\ee
Since $ x^{-2/\alpha} S^6 \rightarrow \infty$, this can go to zero only if
\be
\frac{\dot{S}}{S}(V^2-1) \rightarrow 2
\ee
and eqn (2.19) then implies
\be
\frac{\dot{V}}{V} \rightarrow \frac{1-5\alpha}{1-\alpha}.
\ee
However, since the last term in eqn (3.50) scales as
\be
L^{(\alpha-1)/(5\alpha -1)} \left[\frac{\dot{S}}{S}(V^2-1) - 2 - \frac{S}{\dot{S}}\right],
\ee
we also need the term in square brackets to go to zero as $L^{(1-\alpha)/(5\alpha -1)}$. This determines the second order terms in the expressions for
$x$ and $S$ but does not impose any further relationship
between $x_0$, $S_0$ and $z_*$.
Note that the metric can be written as
\be
ds^2 \sim L^{4\alpha/(1-5\alpha)} [dt^2 - dr^2 -r^2 L^{2(3\alpha -1)/(5\alpha-1)}d\Omega^2]
\ee
This resembles the open Friedmann model and can be transformed to 
Minkowski form with a new choice of time-slicing (as in the Milne model). These solutions are therefore asymptotically flat or, more precisely, asymptotically 
Schwarzschild (since the mass tends to 
a non-zero constant). They are discussed further by Carr et al. (1998).

\vskip .3in
$\bullet$ $V\rightarrow \infty$ as $z\rightarrow z_*$. In this case, eqn (2.16) can be written in the form
\be
aL^{-1}=2\alpha b L^{-1} - \alpha (1+\alpha) \beta^{2} z_*^{-2(1-\alpha)/(1+\alpha)} (x_0 L)^{a(\alpha -1 )/\alpha}.
\ee
It is easy to show that the only consistent solution has $a = \alpha/(1-\alpha)$, so that all the terms scale as $L^{-1}$, and eqn (3.44) then implies $b=2/[3(1-\alpha)]$, so we have
\be
S = S_0 L^{2/3(1-\alpha)},\;\;\;x = x_0 L^{\alpha/(1-\alpha)}.
\ee
Eqn (2.17) then gives 
\be
M \sim L^{-2/3(1-\alpha)},
\ee
so $M \rightarrow \infty$ and $MS$ tends to a constant at $z=z_*$. Note that eqn (3.55) yields the same relation between A, B and $z_*$ as eqns (2.17) and (2.18), so these solutions are described by two
independent parameters. Eqns (2.13) and (2.14) imply that the metric tends to
\be
ds^2 \sim L^{2\alpha/(1-\alpha)}dt^2 - L^{-2/3(1-\alpha)}dr^2 -
r^2 L^{4/3(1-\alpha)}d\Omega^2,
\ee
corresponding to a Schwarzschild-type singularity of infinite density.

\setcounter{equation}{0}
\section{Self-Similar Solutions with Dust}

We first discuss the dust similarity solutions because these exhibit many of the features of the similarity solutions with pressure, even though the
equations are considerably simplified. In particular, if $\alpha=0$, there is no sonic point and  both the mass and energy within comoving radius $r$ are
conserved, so $E$ and $m/r=MS$ are constant. If
we put $m/r=\kappa$, then eqn (2.4) implies
\be
\label{density}
 4\pi\mu R^{2}\frac{\partial R}{\partial r}=\frac{dm}{dr}=\kappa
\ee
and this can be combined with eqns (2.12) and 
(2.14) to give
\be
\label{ssgrr}
   e^{\lambda}=\gamma^{-1}(4\pi\mu r^{2}S^{2})^{-1} =
   \kappa^{-1}\gamma^{-1}\frac{\partial R}{\partial r}.
\ee
On the other hand, the definition of $\Gamma$ in eqn 
(2.6) implies
$e^{\lambda}=\Gamma^{-1}(\partial R/\partial r)$ and so the 
constant $\kappa$
is just $\Gamma/\gamma$. The mass function is therefore
\be
\label{ssmass}
   M=\frac{\Gamma}{\gamma S}=\frac{\sqrt{1+2E}}{\gamma S},
\ee
where we have taken the positive square root to ensure that the mass if positive. (We discuss the negative mass case later.) Since eqn (2.13) implies $e^{\nu}=\beta$,
eqn (2.6) can now be rewritten as 
\be
\label{ssenergy}
   E=\frac{1}{2\beta^{2}}z^{4}\left[\frac{dS}{dz}\right]^2
   -\frac{\sqrt{1+2E}}{\gamma S}.
\ee
This equation is equivalent to eqns (2.17) and (2.18) with $\alpha =0$.
Indeed, from eqn (2.12), providing one makes the substitution
\be
x \rightarrow 1, \;\;\; \frac{1}{\alpha}\frac{\dot{x}}{x} \rightarrow -\frac{dln(\mu r^2)}{dlnz},
\ee
eqns (2.15) to (2.18) with $\alpha =0$ are all formally satisfied. 

In this case, rather than imposing eqn (3.10), it is convenient to 
scale the $r$ and $t$ coordinates so that $\beta=\gamma=1$. This is 
equivalent to making the rescaling $r\rightarrow \gamma r$ and 
$t\rightarrow t/\beta$. Eqn (4.4) then implies 
\be
\label{ssdSdz}
   \frac{dS}{dz} = \pm \frac{\sqrt{2E+2\Gamma/S}}{z^{2}}
\ee
and this can be integrated to give 
\be
\label{ssdust}
   D\pm \frac{1}{z}=\left\{ \begin{array}{ll}

   \frac{\sqrt{ES^{2}+\Gamma S}}{\sqrt{2}E}-
\frac{2\Gamma}{(2E)^{3/2}} 
   \sinh^{-1}\sqrt{\frac{ES}{\Gamma}} & (E>0) \\ & \\

   \frac{\sqrt{2}}{3}S^{3/2} & (E=0) \\ & \\

   \frac{\sqrt{ES^{2}+\Gamma 
S}}{\sqrt{2}E} \pm\frac{2\Gamma}{(-2E)^{3/2}}
   \sin^{-1}\sqrt{\frac{-ES}{\Gamma}} & (-1/2<E<0)

   \end{array} \right.
\ee
where D is an integration constant, the plus and minus signs apply for $dS/dz$
negative and positive, respectively, and $\sin^{-1}$ is always taken to be positive. If we took the negative square root in eqn (4.3),
corresponding to M and $\Gamma$ being negative, there would be another
solution for $E>0$ given by
\be
D\pm \frac{1}{z}= \frac{\sqrt{ES^{2}-|\Gamma| S}}{\sqrt{2}E}+
\frac{2|\Gamma|}{(2E)^{3/2}} 
   \cosh^{-1}\sqrt{\frac{ES}{|\Gamma|}} \;\;\; (E>0). 
\ee
This solution is unphysical, since the mass is negative, but it is of interest for later comparison with the solutions with pressure. Eqns (2.10) and (2.11) give the velocity functions as
\be
\label{ssvel}
   V=\frac{Sz\pm\sqrt{2E+2\Gamma /S}}{\Gamma}
\ee
and
\be
\label{ssvelR}
V_R= \pm\frac{\sqrt{2E+2\Gamma /S}}{\Gamma},
\ee
while eqns (2.10) and (4.2) imply that the density is given by
\be
4\pi \mu t^2 = \frac{1}{zS^2V} = \frac{\sqrt{1+2E}}{zS^2(Sz\pm\sqrt{2E+2\Gamma /S})}
\ee
where the plus and minus signs again correspond to $dS/dz$ being negative and positive, respectively. Note that V and $\mu$ are negative (corresponding to tachyonic models) for the solution given by eqn (4.8).

Eqn (4.7) implies that there is a
2-parameter family of similarity solutions (as in the general 
$\alpha$ case). The flat Friedmann solution has $D=E=0$, so we 
need to determine the significance of the solutions in which $D$ and 
$E$ are non-zero. In each case, we will show the form of $S$, $V$
and $\mu t^2$ as functions of $z$. In obtaining the full family of solutions, it is crucial that we allow $z$ to be either positive or negative. Our analysis will also cover the unphysical solutions with negative mass.

\subsection{$D=0$ solutions}

Solutions with $D=0$ are asymptotically Friedmann as 
$|z|\rightarrow \infty$ and are specified entirely by the energy 
parameter $E$. The forms of $S(z)$ and $V(z)$ in these solutions are shown in Figures (2a) and (2b). The solutions with $z>0$ correspond to initially
expanding Big Bang models and have $V>0$ everywhere: they start from an initial Big Bang singularity ($S=0$) at $t=0$ ($z=\infty$) and then either expand indefinitely ($S\rightarrow \infty$) as $t \rightarrow \infty$ ($z \rightarrow 0$) for $E\ge 0$ or recollapse to a black hole singularity ($S=0$) at
\be
z_S=\frac{(-2E)^{3/2}}{2\pi\sqrt{1+2E}}
\ee
for $E<0$. In the first case, V decreases monotonically from $\infty$ to 0. In the second case, it reaches a minimum before rising to $\infty$ at $ z_S$; the minimum will exceed 1 (in which case the whole Universe is inside the black hole) if $E$ is less than some critical negative value $E_*$ and it will be less than 1 (in which case there is a black hole event horizon and a cosmological particle horizon) if $E$ exceeds $E_*$. Such solutions were studied in detail by Carr \& Hawking (1974) and Carr \& Yahil (1990). The solutions with $z<0$ are the time-reverse of these and have $V<0$ everywhere: 
as $t$ increases from $-\infty$ to 0 (i.e. as $z$ decreases from 0 to $-\infty$),
the $E\ge 0$ models collapse from an infinitely dispersed state ($S=\infty$) to a Big Crunch singularity ($S=0$); the $E<0$ models also collapse to a Big Crunch singularity but they emerge from a white hole and are never infinitely dispersed.

Both $S$ and $V$ have the same $z$-dependence as in the $E=0$ Friedmann solution as $|z|\rightarrow\infty$:
\be
S = [9\sqrt{1+2E}/2]^{1/3}|z|^{-2/3},\;\;\;
V=[6(1+2E)]^{-1/3}z^{1/3}.\label{eq:  } 			
\ee
However, the $E\neq 0$ solutions deviate from the $E=0$ solution at small values of $|z|$. The $E<0$ solutions never reach $z=0$ at all, while the $E>0$ ones have 
\be
S = (2E)^{1/2}|z|^{-1} , \;\;\;
V = -(1+2E)^{1/2} E^{-1}z\,ln[(2E)^{3/2}(1+2E)^{1/2}|z|].\label{eq:  }
\ee
as $|z| \rightarrow 0$. The first relation implies that the circumference function $R(r,t)=Sr$ is non-zero in limit $r\rightarrow 0$ unless $E=0$ since
\be
R(t,0) = \sqrt{2E}\, t.\label{eq:  }
\ee
This means that the ``coordinate'' origin ($r=0$) is an 
expanding 2-sphere. This has a natural physical interpretation since the forms of $S$ and $V$ are similar to those in the KS solution [cf. eqns (3.19) and (3.23)], in which all the matter is localized on a shell. [However, there is no exact KS solution in the dust case.] To obtain a complete solution, one must match the self-similar solution onto a (non-self-similar) void inside $R(t,0)$. In the $E<0$ case, the physical origin is the black hole singularity (since $R=rS=0$ there), so only in the $E=0$ case can one identify $z=0$ with the physical origin.

The form of the density function $\mu t^2$ is shown in Figure (2c). At a given time this specifies the density profile $\mu(r)$ and it illustrates that a non-zero value of $E$ necessarily introduces 
inhomogeneity into the model. Solutions with $E>0$ are everywhere underdense relative to the Friedmann model, whereas those with $E<0$ 
are everywhere overdense. For the latter, the density diverges at the singularity and 
eqns (4.3) and (4.12) imply that the mass associated with this singularity is
\be
m_S=(MSz)_S\;t = (-2E)^{3/2}t /(2\pi).
\ee
It therefore starts off zero when the singularity first forms at $t=0$ but then grows as $t$.

The form of the mass function $M(z)$ 
in the $D=0$ solutions is not shown explicitly but can be immediately deduced from the expression for $S$ since eqn (4.3) implies $M \sim S^{-1}$. 
In the $E\geq 0$ case, there is always a single point where $M=1/2$ and this corresponds to the cosmological apparent horizon. In the $E<0$ case, eqn (4.7) implies that S has a maximum of $\Gamma/|E|$ and so eqn (4.3) shows that 
M has a minimum 
of $|E|$. Since this is less than 1/2, there are always two points where
$M=1/2$, one corresponding to the 
black hole apparent horizon and the other to the cosmological apparent horizon.
Note that a black hole's apparent horizon always lies within or coincides with its event horizon,
which is why the first can exist without the second.

Finally, we consider the $E>0$ negative-mass solutions given by eqn (4.8). Their form is indicated by the broken curves in Figures (2). $S$, $V$ and $\mu$ have the same form as in the positive mass solutions for small values of $|z|$ except that V and $\mu$ reverse their signs.  However, the solutions are very different at large values
of $|z|$ since eqn (4.6) shows that $S$ must always exceed $|\Gamma|/E$. Indeed it tends to this value asymptotically, so we have
\be
S\approx \sqrt{1+2E}/E,\;\;\;V \approx -z/E,\;\;\;\mu r^2 \approx -\frac{E^3}{\sqrt{1+2E}}
\ee
as $|z| \rightarrow \infty$. The form of this solution is closely related to that of the $\alpha <<1$ static solution [cf. eqn (3.29)], although there is no static solution in the $\alpha=0$ case itself. 

\subsection{$E=0$ solutions}

We now put $E=0$ and consider the effect of introducing a non-zero value for the constant $D$. [In this case, eqn (4.6) does not permit $\Gamma <0$, so there are no negative-mass solutions.] The form of $S(z)$ for the $D>0$ solutions is shown in Figure (3a). There are two types of solutions in this case, one expanding and the other collapsing. For the expanding solutions (solid lines), $S=0$ at $z=-1/D$ and so the Big Bang occurs before $t=0$ (i.e. it is ``advanced''). As $t$ increases to 0 (i.e. as $z$ decreases
to $-\infty$), $S$ tends to the finite value
\be
S_\infty(D)=(3D/\sqrt{2})^{2/3}.
\ee
As $t$ further increases from 0 to $+\infty$ (i.e. as $z$ jumps to 
$+\infty$ and then decreases to 0), $S$ increases monotonically to $\infty$. For the contracting solutions (broken lines), $S$ starts infinite at $t=-\infty$ ($z=0$) and then decreases to $S_\infty(D)$ as $t$ increases to 0 ($z \rightarrow -\infty$). As $t$ further increases (i.e. as $z$ jumps to $+\infty$ and then decreases), $S$ continues to decrease until it reaches 0 at the Big Crunch singularity at $z=1/D$. Both types of solutions are characterized by the fact that they have just one singularity. 

The form of $V(z)$ in the $D>0$ solutions is 
shown in Figure (3b). For the expanding solutions (solid lines), it starts off at $-\infty$ 
at the Big Bang ($z=-1/D$), reaches a negative maximum (which will be more than -1 if $D$ exceeds some critical value $D_*$) and then, from eqn (4.9), tends to 
\be
V = S_\infty(D) z = (3D/\sqrt{2})^{2/3} z
\ee
as $z\rightarrow -\infty$. When z jumps $+\infty$, $V$ becomes positive but eqn (4.18) still applies. For large $|z|$, $V \sim z$ rather than $z^{1/3}$ 
(as applies in the $D=0$ case) because any solution with finite $S$ at infinity must be ``nearly'' static in the sense that $dS/dz$ tends to zero. However, the solutions are not asymptotic to an {\it exact} static solution (indeed this does not exist in the $\alpha=0$ case) because eqn (4.10) implies that $V_R$ tends to a non-zero value:
\be
V_R^{\infty} = \left(\frac{4}{3D}\right)^{1/3}.
\ee
This is also reflected by the fact that, in terms of the variable $\xi \equiv 1/z$, $dS/d\xi \neq 0$ at $\xi=0$. We therefore term these solutions asymptotically ``quasi-static''. As $z$
decreases from $+\infty$ to 0, $V$ decreases monotonically to 0. For the contracting models (broken lines), $V$ starts from zero at
$z=0$ and monotonically decreases as $z$ decreases to $-\infty$, being given by eqn (4.19) asymptotically. When $z$ jumps to $+\infty$, $V$ jumps to $+\infty$ and  then decreases to a minimum before rising to infinity  at the Big Crunch singularity. The minimum will be less than 1 if $D$ exceeds the value $D_*$; in this case, one necessarily has a naked singularity.

The form of the density function $\mu t^2$ in the $D>0$ solutions is interesting. From eqn (4.11) the density parameter has the form
\be
\Omega = 6\pi\mu t^2 = \frac{1}{(1\pm 3D|z|)(1\pm D|z|)}
\ee
where the + and - signs apply for negative and positive values of $dS/dz$, respectively. (The factor $6\pi t^2$ corresponds to the density in a flat Friedmann dust universe.) For a given fluid element, this just describes how the density evolves as a function of time and it has the expected form. However, at a given time it also prescribes the density profile and one sees immediately that a non-zero value of $D$ (like a non-zero value of $E$) introduces an inhomogeneity.  This inhomogeneity has a particularly interesting form. In the $z<0$ regime, the profile for the collapsing solutions is homogeneous for $|z|<<1/D$ but roughly static and isothermal ($\mu \sim r^{-2}$) for $|z|>>1/D$. In the $z>0$ regime, the profile corresponds to a density singularity at the centre of a nearly static isothermal sphere. These features are illustrated by the broken lines in Figure 3(c) and have an obvious physical interpretation. It is interesting that the isothermal model features prominentl!
y in 
both regimes despite the fact that there is no exact static solution in the dust case. The mass of the singularity in these solutions is
\be
m_S=(MSz)_S\;t = t /D
\ee
from eqn (4.3). As in the asymptotically Friedmann case, it therefore starts off zero at $t=0$ but then grows as $t$. As pointed out by Ori \& Piran (1990), such solutions have an analogue in Newtonian theory (Larson 1969, Penston 1969).

Finally, we consider the $D<0$ solutions. The form of $S(z)$ in this case is shown by the dotted curves in Figure (3a). Such solutions are confined to $|z|<1/D$, with $S$ either decreasing monotonically for $z<0$ (i.e. as $t$ increases from $-\infty$) or increasing monotonically for $z>0$ (i.e. as $t$ increases to $+\infty$). However, these solutions break down before $S$ reaches 0. This is because, as indicated by the dotted curve in Figure (3b), $|V|$ increases to some maximum value and then falls to zero at $|z|=1/(3D)$. From eqn (4.20) this means that the density diverges there, as indicated by the dotted curve in Figure (3c). This divergence is associated with the formation of a shell-crossing singularity since the model resembles the KS solution at this point. For $|z|>-1/(3D)$, the density and velocity function become negative but this is presumably unphysical. 

\subsection{$D\neq 0, E\neq 0$ solutions}

The forms of S(z) and V(z) for the $(D>0, E\neq0)$ solutions are indicated by the solid and broken lines in Figures (4a) and (4b). The figures assume that $D$ is fixed but
allow $E$ to vary. The $(D>0, E>0)$ solutions are qualitatively similar to the $(D> 0, E=0)$ ones (i.e. one has monotonically expanding or collapsing models). In particular, the form of the solution near $z=0$ is still given by eqns (4.14), so the value of $D$ is unimportant here. As before, $z=0$ is no longer the origin, so one has to attach the solution to a non-self-similar central region. As $|z|\rightarrow\infty$, the solutions with $E>0$ have 
\be
S \approx S_{\infty}(D,E),\;\;\; V \approx S_{\infty}(D,E)(1+2E)^{-1/2} z \label{eq:  } 
\ee
where $S_{\infty}(D,E)$ is the solution of eqn (4.7) with $z=\infty$. As in the  $(D> 0, E=0)$ case, these solutions are only quasi-static and not exactly static at large $|z|$ since eqn (4.10) implies
\be
V_R(\infty) = \Gamma ^{-1}\sqrt{2E + 2\Gamma /S_{\infty}}
\ee
and this is
always positive. Also as in the ($D>0, E=0$) case, $|V|$ reaches a minimum before diverging at the Big Bang or Big Crunch singularity and this minimum will fall below 1, corresponding to a naked singularity, providing $E$ exceeds some critical negative value $E_*(D)$. The dependence of $E_*$ on $D$ can only be determined numerically but it clear that this corresponds to lying outside some boundary in the $(E,D)$ diagram.

The $(D> 0, E<0)$ solutions are qualitatively different
from the $(D> 0, E=0$) solutions in that there are no models which collapse from or expand to infinity. This is clear
from eqn (4.6) which implies that $S$ never exceeds $\Gamma/|E|$. One might expect  that there could be an exact static solution
with $S=\Gamma/(-E)$ but this is incompatible with eqn (2.15) if $\alpha =0$. The models which expand from the Big Bang singularity at $z=-1/D$ recollapse to a black hole singularity at 
\be
z_S=\left[D+\frac{2\pi\sqrt{1+2E}}{(-2E)^{3/2}}\right]^{-1}
\ee
(necessarily less than $1/D$), whereas the models which recollapse to the Big Crunch singularity at $z=1/D$ emerge from a white hole singularity at $-z_S$. Again there will be a naked singularity if the
minimum value of $|V|$ is less than 1. 

The form of $\mu t^2$ in these solutions is shown in Figure (4c). It is very similar to the form indicated in Figure (3c) except that one has added curves corresponding to the $E<0$ solutions. The $E>0$ solutions are everywhere underdense relative to the $E=0$ solutions, whereas the $E<0$ solutions are everywhere overdense. The form of $M(z)$ can be deduced immediately from Figure (3a) and eqn (4.3). As $|z|\rightarrow\infty$, $M 
\rightarrow \sqrt{1+2E}/S_\infty(D,E)$ and this exceeds 1/2 for sufficiently low values of $D$. In this case there is no black hole or cosmological apparent horizon, which is different from the $D=0$ case. 

The $D<0$ solutions have the same form as in the $E=0$ case, except that 
the $E<0$ solutions emerge from a white hole rather than collapsing from  infinity. As in the $E=0$ case, such models are probably physically unrealistic since the density diverges due to shell-crossing. They are therefore not shown explicitly.
The form of the (unphysical) negative-mass solutions, which only exist for $E>0$, is indicated by the dotted curve in the Figures (4) and is similar to the $D=0$ case shown in Figure (2). 

\setcounter{equation}{0}
\section{Self-Similar Solutions with Pressure}

One might expect the $\alpha \neq 0$ solutions to share some of the 
qualitative features as the $\alpha=0$ ones in the 
supersonic regime. The arguments for this are partly physical (viz. 
pressure effects should be unimportant on sufficiently large scales) and 
partly mathematical (viz. the dust equations can be obtained from the 
general $\alpha$ equations by taking the limit $\alpha \rightarrow 0$). The behaviour of the self-similar dust solutions therefore suggests that there should exist at least two classes of similarity solutions with pressure at large $|z|$: a 1-parameter family which are asymptotically Friedmann and a 2-parameter family which are asymptotically quasi-static.
In this section we will show that there are indeed solutions of these kinds. However, we saw in Section 3 that new possible behaviours arise at large $|z|$ when there is pressure. In particular, there is an exact static solution and an exact
Kantowski-Sachs solution, so one might expect there to be families of solutions asymptotic to these. We will demonstrate that this is indeed the case and that each of the families is
described by one parameter. We saw in Section 3 that there are also asymptotically (but not exact) Minkowski solutions for $\alpha >1/5$. We discuss
these solutions in more detail elsewhere (Carr et al. 1998) but do not consider them 
further here. 

The inclusion of pressure also introduces qualitatively new features in the {\it subsonic} regime, so there are important differences from the dust solutions at small values of $|z|$. In particular, the presence of a sonic point at $|V|=\sqrt\alpha$ allows solutions to be discontinous there, so one might anticipate a wide variety of subsonic behaviours.  However,
the requirement that the solution be regular at the sonic point severely 
restricts the possible subsonic behaviours, with the consequence that 
the only possible solutions with pressure at small $|z|$ are the exact static model, a 1-parameter family of asymptotically Friedmann models and a 1-parameter family of asymptotically Kantowski-Sachs models. Indeed the value of $z$ at the sonic point
almost uniqely (modulo the number of oscillations) identifies the subsonic solution, which means that one can effectively extend a solution in the supersonic regime into the subsonic regime in only one way. Although other extensions exist, they will become unphysical at some point (either because the mass
becomes negative or because they encounter another sonic point at which they become irregular).

We will start by discussing the characteristics of the asymptotically Friedmann and asymptotically KS solutions. In this context we will be mainly reviewing the work of Carr \& Yahil (1990) and Carr \& Koutras (1993)
but we will extend these earlier studies somewhat, so as to make the connection with the dust solutions more explicit. We will then discuss the asymptotically 
quasi-static solutions. The discussion here will be entirely original, although some 
examples of this type of solution have been considerd before. The form of the function $V(z)$ for the various solutions 
is shown in Figures 6-8. The $\alpha=1/3$, $\alpha>1/3$ and $\alpha<1/3$ cases are shown separately, since
they have somewhat different characteristics. The $\alpha=1/3$ case is special
since the equations simplify, so this is the one which has been most studied numerically. To 
avoid an unnecessary proliferation of detail, we only present the $z>0$ curves. The $z<0$ curves are trivially derived from these by reflecting in the origin (cf. Figures 2-4): one merely replaces $z$ by $|z|$ and $V$ by $-V$.  However, it is important to stress that, as in the dust case, the
complete solutions may involve both the $z<0$ and $z>0$ regimes.
 
\subsection{Asymptotically Friedmann Solutions}

Carr \& Yahil (1990) consider solutions which are either 
exactly or asymptotically Friedmann for large and small values of $z$. They therefore introduce functions $A(z)$ and $B(z)$ defined by
\be
   x \equiv z^{-2 \alpha/(1+\alpha)} e^A,\;\;\;
   S\equiv z^{-2/3(1+\alpha)} e^B. 
\ee
The Friedmann solution itself ($A=B=0$) passes through the sonic line Q at
\be
 z_F = \left[\frac{\sqrt{6\alpha}}{1+3\alpha}\right]^{3(1+\alpha)/(1+3\alpha)}.
\ee
For comparison the static solution passes through Q at
\be
z_S = \left[\frac{(2\alpha)^{3/2}3^{(5\alpha-1)/2(\alpha-1)}}
{(1+3\alpha)(1+6\alpha + \alpha ^2)}\right]^{(1+\alpha)/(1+3\alpha)}
\ee
and one can show that this is always less than $z_F$. In fact, the only physical subsonic solution which passes through $z_F$ is the Friedmann solution itself. All the other solutions are unphysical because, as $z$ decreases, $V$ either reaches a minimum and then hits the sonic surface again but off the line $Q$ or the mass within $r$ goes negative (Bicknell \& Henriksen 1978a). Since the only solution which is exactly Friedmann outside the sonic point is the Friedmann solution {\it everywhere}, we confine attention to solutions which are asymptotically Friedmann at large $z$. 

The ordinary differential equations for $x$ and $S$ now become ordinary
differential equations for $A$ and $B$.
If we linearize the equations in $A$, $\dot{A}$ and $\dot{B}$ to find the 1st order solution as $z \rightarrow \infty$, eqns (2.15) and (2.16) yield
\be
\ddot{B} = \left(\frac{1}{3\alpha}\right)\dot{A} - \left(\frac{1+3\alpha}{1+\alpha}\right)\dot{B},
\ee
\be
\dot{B} = \left(\frac{1}{2\alpha}\right)\dot{A} - \frac{(\alpha-1)}{3\alpha(1+\alpha)}A\;.
\ee
Differentiating the second equation and eliminating $\dot{B}$ and $\ddot{B}$ then leads to the following differential equation for $A$:
\be
\ddot{A}+\frac{(9\alpha-1)}{3(1+\alpha)}\dot{A}+
\frac{2(1+3\alpha)(\alpha-1)}{3(1+\alpha)^2}A =0.
\ee
This has two solutions,
\be
A \sim z^{-2(1+3\alpha)/3(1+\alpha)}\;\;\;
or \;\;\; A \sim z^{(1-\alpha)/(1+\alpha)},
\ee
but the second one can be rejected since the exponent is positive for $\alpha<1$ (so that $A$ diverges as $z\rightarrow \infty$). The first solution
then gives
\be
   A = - \frac{\alpha (1 + 3 \alpha)}{(1+ \alpha)} kz ^{-2(1+3 \alpha)/3(1+\alpha)},\;\;\;
   B = B_\infty -kz^{-2(1+3\alpha)/3(1+\alpha)}
\ee
where $B_\infty$ and $k$ are integration constants.  Note that $A \rightarrow 0$ as $z \rightarrow \infty$ because 
eqns (2.12) and (5.1) show that $A$ has physical significance (viz. the density
perturbation) but there is no physical restriction on $B_\infty$ (the asymptotic
value of $B$). The constants in eqn (5.8) are related since eqns (2.17) and (2.18) imply 
\be
   k = \frac{3(1+\alpha)(e^{-2B_\infty} -e^{4B_\infty})}{2(1+3 \alpha)
  (5+3  \alpha)} .
\ee
Thus there is a $1$-parameter family of asymptotically Friedmann solutions. 
[Despite the presence of the parameter $B_\infty$, these solutions are really asymptotic to the {\it exact} Friedmann model, since one could formally gauge $B_\infty$ to zero at infinity (but not finite $z$) by taking a different spatial
hypersurface.]
>From eqn (2.20), the energy function is 
\be
E = E_\infty + O(z^{-4(1+3\alpha)/3(1+\alpha)}), \;\;\;
E_\infty = \frac{1}{2}(e^{6B_{\infty}}-1)
\ee
Eqn (2.12) shows that the solutions are overdense or underdense relative to the Friedmann solution according to whether $A < 0$ 
or $A > 0$, respectively. From eqns (5.9) and (5.10), this corresponds to $(k >0, B_\infty < 0, E_\infty <0)$ or $(k <0, B_\infty > 0, E_\infty > 0)$, respectively. In the Friedmann case itself, $k = B_\infty = E_\infty = 0$.

If $B_\infty$ is sufficiently  negative, $V$ reaches a minimum value $V_{\min}$ above $\sqrt{\alpha}$ as $z$ decreases and then rises again to infinity.  Such solutions
are supersonic everywhere and contain black holes which grow as fast as
the Universe. There is an event horizon and particle horizon 
providing $V_{\min} <1$ and this will apply if $B_\infty$ is more than some critical negative value $B_\infty^*$; otherwise the whole Universe
is inside the black hole. However, since eqn (2.17) implies that $\dot{M}=0$ at 
$\dot{S}=0$ and eqn (2.18) then implies $M<1/2$, there is still an {\it apparent} horizon for $B_\infty < B_\infty^*$.; this generalizes the result found in the
dust case. As $B_\infty$ increases (i.e. as the asymptotic 
overdensity decreases), the values of $V_{\min}$ and $z_{\min}$ decrease.
Eventually it reaches another critical negative value $B_\infty^{crit}$ at which $V_{min} = \sqrt{\alpha}$ and, for 
$B_\infty > B_\infty^{crit}$, the solutions must reach the sonic surface. 
>From eqn (5.10), these critical negative values for $B_\infty$ also correspond to critical negative values for the asymptotic energy, $E_*$ and $E_{crit}$.  As $B_\infty$ continues
to increase, the value of $z$ at which the solution goes transonic ($z_s$)
increases, passing through the value indicated by eqn (5.2) when $B_\infty =0$ 
and tending to infinity as $B_\infty$ goes to infinity
(corresponding to increasingly underdense solutions). All the solutions
with $B_\infty > B_\infty^{crit}$ reach the sonic surface but only the ones
which cut it on the line $Q$ are regular. This only applies if
$z_s$ lies within the ranges $z_1$ to $z_2$ or above $z_3$ [as indicated in Figure (1)]. $\dot{A}$ diverges at the sonic point for values of $B_\infty$ corresponding to $z_2 < z_s <z_3$. 

We next consider the condition that the solution be well-behaved at 
$z =0$, in the sense that the density and velocity are finite.  This
requires that $A$ and $B$ be finite, which implies
\be
V(0)=0,\;\;\;\dot{A}(0) = \dot{B}(0). 
\ee
We also require $M(0) =0$, and eqn (2.18) then implies 
\be
A(0) = 3 \alpha B(0), 
\ee
which shows that there is a $1$-parameter family of solutions which are
well-behaved at the origin. We will take this parameter to be $A_0 \equiv A(0)$. This is a measure of the overdensity at the origin since, from eqns (2.12) and (5.1),
\be
A_0 = \left[\frac{\alpha}{1 + \alpha}  \right] \log \left[  \frac{\mu_F(0)}{\mu(0)}\right] 
\ee
where $\mu_F$ is the density in the Friedmann solution. Thus $A_0 > 0$
and $A_0 < 0$ solutions are underdense and overdense, respectively. Eqn (2.6) implies that $E\rightarrow 0$ as $z \rightarrow 0$, so the energy function goes to zero as $S\rightarrow \infty$ (as expected).
 
For some range of values of $A_0$ the subsonic solutions must hit the sonic
surface in $(x, S, \dot{S})$ space, since the solution with $A_0 =0$ does.  
As before, the regular solutions must hit it on the line $Q$. However, the behaviour of the subsonic parts of the solutions is more
complicated than that of the supersonic solutions.  As the parameter $A_0$ decreases from positive values to some 
critical negative value $A_0^{crit}$, $z_s$ decreases continuously to $z_1$. In this parameter range the solutions with $z_s > z_3$ are regular at the sonic 
point, while those with $z_1 < z_s < z_3$ are all irregular. [In the $\alpha =1/3$ case, for which $z_3=z_F$, one therefore has a
continuous family of underdense solutions but no overdense
solutions within this range.]  As $A_0$ decreases below $A_0^{crit}$, the $V(z)$ curves develop an inflexion and $z_s$ increases again to the value $z_2$.  The subsonic solutions thus cross over each other in $V(z)$ space.  Although 
one does not reach every value of $z_s$ between $z_1$ and $z_2$, there is a
band of solutions within $z_1 < z < z_2$ which are regular at the sonic point.  This corresponds to the first band of overdense solutions and is 
associated with just a small range of $A_0$ values. As $A_0$ decreases further, $z_s$ moves back and forth between the values $z_1$ and  $z_2$ and the 
$V(z)$ curves exhibit an increasing number of oscillations.  One can group
these solutions into families according to the number of oscillations they
exhibit. Each family will contain a narrow band of solutions (corresponding 
to bands of $A_0$ values) which are regular at the sonic point. The existence of these bands was first pointed out by Ori \& Piran (1990). Indeed this band structure arises even in the Newtonian situation, as shown by Whitworth \& Summers (1985). Note that all the overdense solutions are
nearly static close to the sonic point (in fact, $z_2 = z_S$ if $\alpha =1/3$), although they
deviate from the static solution as one goes towards the origin.

Provided there are points on $Q$ which are intersected by solutions which 
are asymptotically
Friedmann at both large and small $z$, one can construct a solution with a sound-wave which represents a density perturbation growing at the same rate as
the Universe.  One would expect this to be possible providing there is a $1$-parameter family of solutions at the sonic
point, i.e. providing $z_s$ lies in the range of values between $z_1$ and $z_2$ and above $z_3$. For in this case, for each point on $Q$, one would expect at least one supersonic solution to be asymptotically Friedmann and at least one
subsonic solution to be regular at the origin. Furthermore, one
would expect the value of $\dot{A}$ (corresponding to the density gradient or velocity gradient) to be continuous at the sonic point in such 
solutions since only one value corresponds to the $1$-parameter family. The numerical calculations of Carr \& Yahil (1990) show that transonic solutions do indeed exist and have the features aniticipated.  In particular, the solutions all have
a continuous velocity gradient at the sonic point, although they do not span
the entire range of values $z_1 < z_s < z_2$. Note that for each $\alpha$ there is also one asymptotically Friedmann solution
which can be attached to the exact static solution inside the sonic point.

The form of the asymptotically Friedmann solutions for $\alpha =1/3$, the case most likely to pertain in the early Universe, is shown in Figure (5).
The qualitative form of the solutions for more more
general values of $\alpha$ is indicated in Figures 6-8. In deriving the form of the velocity function, we use eqns (2.19), (5.1), (5.8) and (5.12) to express the
asymptotic behaviour as
\be
V \approx e^{(1-3\alpha)A_o/\alpha}V_F \quad (|z| << 1), \quad 
V \approx e^{-2B_\infty}V_F  \quad (|z| >> 1)
\ee
where $V_F$ is the exact Friedmann velocity. For $\alpha >1/3$, $V$ is more (less) than $V_F$ for the overdense (underdense) solutions at all values of $z$. However,
for $\alpha <1/3$, $V$ starts below (above) $V_F$ at small $z$ and ends up above
(below) it at large z for the overdense (underdense) solutions. For 
$\alpha =1/3$, the dependence of $V$ upon $A_o$ at small $|z|$ only appears at second order.

\subsection {Asymptotically Kantowski-Sachs Solutions}

If we wish to consider solutions which are asymptotic to the self-similar KS model, then eqn (3.19) suggests that we introduce functions $A(z)$ and $B(z)$ defined by
\be
x = x_0 z^{-2\alpha/(1+ \alpha)}e^A \quad, \quad S = S_0 z^{-1} e^B. 
\ee
As in the asymptotically Friedmann case, we linearize the equations in $A$, $\dot{A}$ and $\dot{B}$ to find the 1st order solution as $|V| \rightarrow \infty$. Eqns (2.15) and (2.16) then yield
\be
\ddot{B} = -\dot{A} + \left(\frac{1+3\alpha}{1+\alpha}\right)\dot{B},
\ee
\be
\dot{B} = \left(\frac{1}{2\alpha}\right)\dot{A} + \left(\frac{1-\alpha}{1+\alpha}\right)A.
\ee
If we differentiate the second equation and then substitute for $\dot{B}$ using the first, one obtains the following differential equation for A:
\be
\ddot{A}+\left(\frac{\alpha-1}{\alpha+1}\right)\dot{A}-
\frac{2\alpha(1+3\alpha)(1-\alpha)}{(1+\alpha)^2}A =0.
\ee
This has two solutions: $A \propto z^{-p_{1}}$ and $A \propto z^{-p_{2}}$ where
\be
p_{1,2} = \frac{-1 + \alpha \pm \sqrt{(1 - \alpha)(24\alpha^2 + 7 \alpha +1)}}{2 (1+\alpha)}.
\ee
For $\alpha > 0$, the KS solution has $|V| \rightarrow \infty$ as $z \rightarrow \infty$, so we must choose the
positive root $p_{1}$. The general solution then has the form
\be
A = A_\infty z^{-p_{1}},\;\;\; B = A_\infty \left[\frac{1}{2 \alpha} - \left(\frac{1 - \alpha}{1 + \alpha}\right) \frac{1}{p_1}\right]z^{-p_{1}} 
\ee
where $A_\infty$ is an integration constant. For $-1< \alpha < - 1/3$, the KS solution has
$|V| \rightarrow \infty$ as $z \rightarrow 0$, so we must choose a negative root. Only $p_{2}$ is negative for this range of $\alpha$, so this gives a solution like eqn (5.20) but with $p_{2}$ replacing $p_{1}$. In both cases there is thus a $1$-parameter family of asymptotically KS solutions.  For $-1/3 < \alpha < 0$, one again has  $|V| \rightarrow \infty$ as $z \rightarrow \infty$ but both $p_{1}$ and $p_{2}$
are negative, so there is no solution as $z\rightarrow \infty$.
[Since eqn (3.31) shows that KS solutions with  $-1/3 < \alpha < 0$ also correspond to static solutions with $0<\alpha<1$ if $r$ and $t$ are interchanged (so that $z$ goes to $1/z$), this is related to the fact - shown later - that there are no asymptotically static solutions as $z \rightarrow 0$ for $0<\alpha<1$.] Eqn (2.20) implies that $E\rightarrow -1/2$ as $z\rightarrow \infty$, so the dust analysis suggests that these are in some sense extreme black hole solutions. 

At small values of $|V|$, $\dot{A} = \dot{B} = 0$ and eqns (2.17) and (2.18) imply that the values of $A$ and $B$ at $z=0$ are related by
\be
e^{2B_0}= \frac{1}{2} [M_{KS} e^{-A_0(1+\alpha)/\alpha} - (M_{KS} -\frac{1}{2}) e^{-2A_0}]^{-1}, 
\ee
where $M_{KS}$ is defined by eqn (3.23). 
Again there is a $1$-parameter family of solutions and $A_0$ is a measure of the underdensity or overdensity at the origin relative to the exact KS solution.  Note that
there are only isolated solutions at a sonic point for $0<\alpha<1$ and this means that any asymptotically KS solution which hits the 
sonic surface is unlikely to be regular there. For although there is a $1$-parameter family of solutions from each point of the line $Q$, corresponding to the freedom in the value of $\dot{A}$, only two of these are
regular and they are both isolated.  It is therefore unlikely that either of them 
will be asymptotically KS since the general solution has two parameters, 
whereas the asymptotically KS solutions have only one.

Carr \& Koutras (1993) have integrated the equations in the $\alpha = 1/3$ case, 
although the physical significance of these solutions is unclear.
Figures (6) show the form of $V(z)$ in this case.  Let us
first consider the supersonic solutions.  The underdense solutions have $A_\infty$ positive.  As $z$ decreases, they all cross $V= -1$ at some point to the left of the exact 
KS solution.  However, they do not hit the sonic point but reach a 
maximum between $V= -1/\sqrt{3}$ and $V = -1$ as $z$ decreases.  They then hit the
$V =-1$ surface again (all with the same value of $z$), with $M$ and $\mu$ 
tending to $0$ and the scale factor $S$ diverging.  [This behaviour is analogous to
that which arises for the solutions which are asymptotically Minkowski at finite $z$.] The overdense
supersonic solutions have $A_\infty$ negative and, as $z$ decreases, they all hit the sonic line to the right of the
exact KS solution.  As $A_\infty$ decreases, the point at which they hit the
sonic line moves to infinity.  All the supersonic solutions have $M<0$ everywhere and so never cross the $M=0$ curve. Let us now consider the subsonic solutions.  The overdense ones have
$A_0$ negative.  None of the curves hit the sonic surface since they reach
a minimum as $z$ decreases and then asymptotically approach $V=0$.  The interesting feature of these solutions is that the function $M$, which is 
negative at the origin, goes through $0$ [i.e. the solution crosses the
curve $M=0$ in $V(z)$ space] and eventually becomes positive as $z$ increases.  $M$ and $\mu$ tend to infinity as $z\rightarrow \infty$.  The underdense solutions 
have $A_0$ positive and hit the sonic line to the left of KS.

\subsection {Asymptotically Static and Quasi-Static Solutions}

If we wish to consider solutions which are asymptotically static, we introduce functions $A(z)$ and $B(z)$ defined by
\be
x = x_o e^A, \;\;\;\ S = S_o e^B 
\ee
where $x_o$ and $S_o$ are given by eqn (3.30). Eqns (2.15) and (2.16) then become
\be
\ddot{B} +3\dot{B}^2 - \frac{\dot{A}}{\alpha} + \left(\frac{\alpha +3}{\alpha +1}\right)\dot{B} - \left(\frac{1+\alpha}{\alpha}\right)\dot{A}\dot{B} = 0
\ee
\be
V^2 \left(\dot{B}- \frac{\dot{A}}{2\alpha}\right) = - \frac{\dot{A}}{2} + 
\left(\frac{\alpha}{1+\alpha}\right)[e^{-4B + A(1-\alpha)/\alpha}-1].
\ee
To find the first order solution as $V \rightarrow \infty$ (i.e. as $z \rightarrow \infty$), we linearize these equations to obtain
\be
\ddot{B} = \frac{\dot{A}}{\alpha} + \left(\frac{\alpha +3}{\alpha +1}\right)\dot{B}
\ee
\be
\dot{B} = \dot{A}/2\alpha
\ee
where the second equation is required since the right-hand-side of eqn (5.24) must be finite as $V \rightarrow \infty$. Eliminating $\dot{A}$ gives
\be
\ddot{B} + \left(\frac{1-\alpha}{1+\alpha}\right) \dot{B} = 0
\ee
and this leads to the general solution
\be
A = A_{\infty} + Cz^{-(1-\alpha)/(1+\alpha)},\;\;\;
B = B_{\infty} + \left(\frac{C}{2\alpha}\right) z^{-(1-\alpha)/(1+\alpha)}
\ee
where $A_{\infty}$, $B_{\infty}$ and C are integration constants. Eqns (2.17) and (2.18) give another relationship between these constants:
\be
C \sim \left[e^{2B_{\infty}-(1+\alpha)A_{\infty}/\alpha} - \left(\frac{1+6\alpha + \alpha^2}{4\alpha}\right) +   \frac{(1+\alpha)^2}{4\alpha}e^{6B_{\infty}-2A_{\infty}/\alpha}\right]^{1/2}e^{A_{\infty} - B_{\infty}}
\ee
where we have omitted a prefactor which depends on $\alpha$. This shows that the asymptotically static solutions are described by two parameters for a given equation of state.

It should be stressed that  the description ``asymptotically static'' in this context is rather misleading. This is because eqns (2.20) and (5.28) imply
\be
V_R \approx - \frac{(1-\alpha)}{2\alpha(1+\alpha)} C
\ee
at large $z$, so only the 1-parameter family of solutions with $C=0$ are asymptotically  static in the sense that the fluid is not moving with respect to the spheres of constant $R$. This agrees with the description of Foglizzo \& Henriksen (1993), who term such solutions ``symmetric''. The {\it exact} static solution has $A_{\infty}=B_{\infty}=C=0$, so the limit of the asymptotically static solutions need not be the exact static solution itself. We will describe the more general solutions with $C\neq 0$ as
asymptotically ``quasi-static'' since they still have $\dot{S}$ and $\dot{x}$ tending to 0 at infinity.

At large values of $z$, eqn (2.20) implies that the energy function in these solutions is given by
\be
E = \frac{(1+\alpha)^2}{2(1+6\alpha +\alpha^2)}e^{6B_\infty - 2A_\infty /\alpha} \left[1-\frac{C(3-\alpha)}{\alpha(1+\alpha)}z^{-(1-\alpha)/(1+\alpha)}
\right] - \frac{1}{2}\;\;.
\ee
where we have used eqns (5.28). The analogue of the parameter $E$ in the dust solutions is therefore
\be
E_\infty \sim e^{6 B_\infty - 2A_\infty/\alpha}.
\ee
In the exact static case, the asymptotic value of $E$ is just $-M$, where $M$ is 
given by eqn (3.29). The analogue
of the parameter $D$ is more complicated but it is related to the value of $V_R$ at infinity. If one considers solutions in which $E\rightarrow 0$ as $z \rightarrow \infty$ and then defines $D$ by analogy with eqn (4.19), eqns (5.30) and (5.31) imply
\be
D \sim e^{3(1-\alpha)A_{\infty}/(2\alpha)}.
\ee
It would be more natural to identify $D$ with the value of $z$ at the singularity (viz. $z_{\infty}=1/D$). However, this value can only be determined numerically
and cannot be expressed in terms of the asymptotic parameters explicitly.

To find asymptotically quasi-static solutions at small values of $z$,
one seeks solutions with $V=0$ and finite values of $A$ and $B$ at $z=0$. However, this requires $\dot{A} = \dot{B} =0$, which from eqn (5.24) implies
\be
4B_o = \left(\frac{1-\alpha}{\alpha}\right) A_o.
\ee
It is easy to see that this condition is incompatible with eqns (2.17)
and (2.18), so there are no asymptotically quasi-static solutions at the origin
(only the exact static solution itself). If instead we seek solutions in which $\dot{A}$ and $\dot{B}$ are finite and non-zero at $z=0$,
so that $A$ and $B$ diverge logarithmically, then eqn (5.24) requires
\be
\dot{A} = -\left(\frac{2\alpha}{1+\alpha}\right).
\ee
Substituting this into eqn (5.23) gives 
\be
3(1+\alpha)\dot{B}^2 + (5+3\alpha)\dot{B} +2 = 0,
\ee
which has the two roots
\be
\dot{B} = -\frac{2}{3(1+\alpha)}\;\;\;or\;\;\;\dot{B}=-1.
\ee
However, these roots just correspond to solutions which are asymptotically Friedmann or asymptotically KS at $z=0$. We analysed such solutions in the 
previous sections, so the asymptotic quasi-static solutions do not yield any new
behaviour at the origin. 

The qualitative features of the asymptotically quasi-static solutions in the $z>0$ regime are indicated in Figures (6) for $\alpha=1/3$, Figures (7) for $\alpha<1/3$ and Figures (8) for $\alpha>1/3$. As in the asymptotically Friedmann case, there are black hole solutions, a continuum of underdense solutions and discrete bands of overdense solutions (although the
oscillations associated with the overdense bands are not exhibited in
the figures). In each case, sample values of $E_\infty$ are shown and this illustrates
the significance of the critical values $E_*$, $E_{crit}$ and $E_{stat}$.

We have already emphasized that the $z<0$ solutions can be obtained from the 
$z>0$ ones by reflection about the origin. As in the dust case, the full family
of solutions involve both regions. Indeed one has a figure
precisely analagous to Figure (4), except that one now has a
sonic point. Thus the collapsing solutions start in 
the $z<0$ region (following the reflection of the curves which go through the origin) and then jump to the $z>0$ region
(following the curves which approach a singularity). Likewise, the expanding solutions start off following the reflections of
the curves which approach the singularity and then jump to the curves which pass through the origin.
The fact that the solutions span both negative and positive $z$ has the
important consequence that there may be more than one sonic point. The collapsing solutions have one sonic point in the $z<0$ regime but may have
another two in the $z>0$ regime if the minimum of $V$ is less than $\sqrt{\alpha}$. Likewise, the expanding solutions may have two sonic points in the $z<0$ regime if the maximum of $V$ exceeds $-\sqrt{\alpha}$ and then another one in the $z>0$ regime. On the other hand, it is unlikely that solutions with more than one sonic point would be regular, so one would expect physical solutions to exist only for values of the parameters such that the maximum of $|V|$ exceeds $\sqrt{\alpha}$. 

A more extensive and quantitative analysis of the asymptotically quasi-static solutions, including a discussion of their regularity at the sonic point, is given elsewhere (Carr et al. 1998). In fact,
some of the 2-parameter family of similarity solutions with pressure
have already been studied numerically by Foglizzo \& Henriksen (1993), although they only focus on the collapsing solutions. [The relationship between their variables and ours is given in Appendix B.] They confirm many of the qualitative features described above. In particular, they show that the solutions are described by two parameters at large $|z|$
and by one parameter at small $|z|$ and they find the expected behaviour
at the sonic point. In their phase space analysis, the orbits corresponding to the overdense solutions converge on and then spiral around the static solution for a while before heading to the origin. This corresponds to the oscillations found by Carr \& Yahil (1990) and Ori \& Piran (1990), with the number of oscillations identifying the overdensity band. Foglizzo \& Henriksen (1993) confirm that the solutions
with $V_{min}<1$ exhibit naked singularities. Indeed, the static attractor is closely related to the critical self-similar solutions found in the collapse calculations of Evans \& Coleman (1995) and Maison (1996). This is discussed in more detail by Carr, Henriksen \& Levy (1998).

\section{Discussion}

In this paper we have analysed the complete family of spherically symmetric
similarity solutions for a perfect fluid with equation of state $p=\alpha \mu$.
The form of $V(z)$ in these solutions is shown explicitly in Figures (6) to (8), although it should be emphasized that the figures are only qualitative
and we have
not included the (subsonic) solutions which are irregular at the sonic point. The
key steps underlying our analysis are: (1) a delineation of the possible asymptotic forms at large and small distances from the origin; (2) an
elucidation of the link between the $z>0$ and $z<0$ solutions; and (3) an explicit derivation of the dust solutions (which can be expressed analytically).

In claiming that our classification is ``complete'', it should be emphasized
that our considerations have been confined to similarity solutions of the first kind (i.e., homothetic solutions in which the similarity variable is $z\equiv r/t$). However, it may be
possible to extend this work to the classification
of similarity solutions of the ``second'' kind. For example, in spherically symmetric perfect fluid
solutions which possess kinematic self-similarity, the similarity variable is of the form
$z=r/t^a$, where the $a$ depends on some dimensional 
constant which contains a scale (Carter \& Henriksen 1989). There
is evidence 
that such solutions asymptote towards exact solutions that admit a homothetic vector 
(Benoit \& Coley 1998), so the asymptotic analysis in this paper may be of rather more general application than is at first apparent;
i.e. the asymptotic behaviour of {\it all} similarity 
solutions (not only those of the first kind) may be determined by the solutions
described in this paper.  Of course, the behaviour at finite values of the similarity variable,
including for example the behaviour at sonic points and horizons, may be quite different.

We have also confined
attention to perfect fluids with a barotropic equation of state (necessarily of the form 
$p = \alpha \mu$) and so our analysis does not cover more general perfect fluids or anisotropic fluids, even though these may be of physical interest.  In
particular, a two-perfect-fluid  model, in which each component is necessarily comoving and has an equation of state of the form $p_i = \alpha_i \mu_i$ ($i = 1, 2$), is formally equivalent to a single perfect fluid that does not have an equation of state.  It is therefore plausible that 
perfect fluid models for which $p/\mu$ is asymptotically constant
may have the same asymptotic behaviour as 
the self-similar solutions studied in this paper. This is indeed 
the case for the two-fluid models in which each component separately satisfies the conservation equations (see Appendix C).

We have not here considered the stiff case ($\alpha =1$), in which the speed of sound is equal to the speed of light. Since
$\alpha =1$ is a bifurcation value, there can be
significant changes in the qualitative behaviour
from the $\alpha <1$ case. Therefore a discussion of stiff perfect
fluid solutions may be important in understanding the dynamics of 
the complete class of similarity solutions.  A partial analysis has
been made by Lin et al. (1976) and Bicknell \&
Henriksen (1978b).  If $\alpha =1$, eqn (2.19) implies that $V$ has no explicit dependence on $z$ and, when $V \neq 1$,  eqns (2.15) and (2.16) yield a single, second-order
autonomous ODE for $S$.  This equation can be better
studied using different mathematical techniques to those employed
in this paper (cf. Carr et al. 1998). In this context, it should be emphasized that our analysis does not cover the case in which the source is a massless scalar field since (if there is no scalar
potential) this is formally equivalent to a stiff fluid whenever the gradient of the scalar
field is timelike. The relevance of self-similar solutions to the occurrence of 
critical phenomena in scalar field collapse has been studied by Choptuik (1993), Abrahams \& Evans 1993, Brady (1994), Gundlach (1995), Koike et al. (1995) and Frolov (1997). When $\alpha =-1$, the perfect fluid source is equivalent
to a cosmological constant.  In this case, a scale is introduced and so there are no self-similar
solutions of the first kind. However, spherically symmetric 
self-similar solutions 
of the more general kind are still possible in this case 
(Henriksen, Emslie \& Wesson 1983).

Two special cases which have not been considered
here are those in which the homothetic vector is either parallel or orthogonal
to the fluid velocity.  These cases are not covered by 
the analysis of Section 2. However,
it can be shown that all perfect fluid spacetimes (not only spherically 
symmetric ones) admitting a homothetic vector
parallel to the velocity vector are necessarily Friedmann
(Coley 1991).  In addition, Ponce de Leon (1993) has claimed that
all spherically symmetric spacetimes which admit a homothetic vector
orthogonal to the velocity vector have a singular metric 
(cf. Bogoyavlensky 1985).

\vskip .5in

{\large \bf Acknowledgments}

We thank Martin Goliath, Dick Henriksen, Ulf Nilsson and Claes Uggla for useful discussions and Andrew Whinnett  for help with the numerical work. An earlier version of this paper omitted the asymptotically Minkowski solutions and we extended our analyis to cover these only after their existence was ascertained by Goliath, Nilsson and Uggla numerically. BJC is grateful to the Department of Mathematics and Statistics at Dalhousie University for hospitality received during this work.

\newpage

\noindent
{\large \bf References}

 \begin{enumerate}

\item[] A. M. Abrahams and C. R. Evans, 1993, Phys. Rev. Lett. {\bf 70}, 2980.

\item[] G. I. Barenblatt and Ya B. Zeldovich, 1972, Ann. Rev. Fluid 
Mech. {\bf 4}, 285.

\item[]  P.M. Benoit and A.A. Coley, 1998, Class. Quantum Grav. {\bf 15}, 2397.

\item[] E. Bertschinger, 1985, Ap. J. {\bf 268}, 17.

\item[] G. V. Bicknell and R. N. Henriksen, 1978a, Ap. J. {\bf 219}, 1043.

\item[] G. V. Bicknell and R. N. Henriksen, 1978b, Ap. J. {\bf 225}, 237.

\item[] G. V. Bicknell and R. N. Henriksen, 1979, Ap. J. {\bf 232}, 670.

\item[] O. I. Bogoyavlenski, 1977, Sov. Phys. JETP {\bf 46}, 634.

\item[] O. I. Bogoyavlenski, 1985, {\it Methods in the Qualitative Theory of  Dynamical Systems in Astrophysics and Gas Dynamics} (Springer-Verlag).

\item[] P. R. Brady, 1994, Class. Quant. Grav. {\bf 11}, 1255.

\item[] A. H. Cahill and M. E. Taub, 1971, Comm. Math. Phys. {\bf 21}, 1.

\item[]B.J.Carr, 1995, preprint prepared for but omitted from {\it The Origin of Structure in the Universe}, ed. E.Gunzig and P.Nardone (Kluwer).

\item[] B. J. Carr, 1998, in {\it Proceedings of the Seventh Canadian Conference on General Relativity and Relativistic Astrophysics}, ed. D. Hobill.

\item[] B. J. Carr and A. A. Coley, 1998a, Class. Quant. Grav., in press.

\item[] B. J. Carr and A. A. Coley, 1998b, Class. Quant. Grav., in press.

\item[] B. J. Carr and S. W. Hawking, 1974, MNRAS {\bf 168}, 399.

\item[] B. J. Carr and A. Koutras, 1993, Ap. J. {\bf 405}, 34 (CK).

\item[] B. J.  Carr and A. Yahil, 1990, Ap. J. {\bf 360}, 330 (CY).

\item[] B. J.  Carr and A. Whinnett, 1998, preprint.

\item[] B. J. Carr, R. N. Henriksen and M. Levy, 1998, preprint.

\item[] B. J. Carr, A. A. Coley, M. Goliath, U.S. Nilsson and C. Uggla, 1998, preprint.

\item[] B. Carter and R. N. Henriksen, 1989, Ann. Phys. Supp. {\bf 14}, 47.

\item[] M. W. Choptuik, 1993, Phys. Rev. Lett. {\bf 70}, 9.

\item[] M. W. Choptuik, 1994, in {\it 
Deterministic Chaos in General Relativity}, ed. D. Hobill et al. 
(Plenum, New York).

\item[]  A. A. Coley, 1991, Class. Quant. Grav. {\bf 8}, 955.

\item[] A. A. Coley, 1997. in {\it Proceedings of the Sixth Canadian Conference on General Relativity and Relativistic Astrophysics}, The Fields Institute 
Communications Series (AMS), Volume 15, page 19, eds. S. P. Braham, J. D. Gegenberg and R. J. McKellar (Providence, R.I.).

\item[] A. A. Coley and B. O. J. Tupper, 1989, J. Math. Phys. {\bf 30}, 2616.
 
\item[] C. B. Collins, 1977, J. Math. Phys. {\bf 18}, 2116.

\item[] D. M. Eardley, 1974, Comm. Math. Phys. {\bf 37}, 287.

\item[] C. R. Evans and J. S. Coleman, 1994, Phys. Rev. Lett. {\bf 72}, 1782.

\item[] T. Foglizzo and R. N. Henriksen, 1992, Phys. Rev. D. {\bf 48}, 4645 (FH).

\item[] A. V. Frolov, 1997, preprint.

\item[] M. Goliath, U. S. Nilsson and C. Uggla, 1998a, Class. Quant. Grav. {\bf 15}, 167. 

\item[] M. Goliath, U. S. Nilsson and C. Uggla, 1998b, Class. Quant. Grav. {\bf 15}, 2841. 
\item[] C. Gundlach, 1995, Phys. Rev. Lett. {\bf 75}, 3214.

\item[] R. S. Hamade, J. H. Horne and J. M. Stewart, 1996, Class. Quant. Grav. {\bf 13}, 2241.

\item[] R. N. Henriksen and K. Patel, 1991, Phys. Red. D. {\bf 42}, 1068.

\item[] R. N. Henriksen and P. S. Wesson, 1978, Astrophys. Space Sci. {\bf 53}, 429.

\item[] S. Ikeuchi, K. Tomisaka and J. Ostriker, 1983, Ap. J. 
{\bf 265}, 583.

\item[] P. S. Joshi and I. H. Dwivedi, 1993, Phys. Rev. D. 47, 5357.

\item[] R. Kantowski and R. Sachs, 1966, J. Math. Phys. {\bf 7}, 443.

\item[] T. Koike, T. Hara and  S. Adachi, 1995, Phys. Rev. Lett. {\bf 74}, 5170.
 
\item[] K. Lake, 1992, Phys. Rev. Lett. {\bf 68}, 3129.

\item[] R. B. Larson, 1969, MNRAS {\bf 145}, 271.

\item[] D. N. C. Lin, B. J. Carr and S. D. M. Fall, 1978, MNRAS {\bf 177}, 151.

\item[] D. Maison, 1996, Phys. Lett. B. 366, 82.

\item[] B. D. Miller, 1976, Ap. J. {\bf 208}, 275.

\item[] C. W. Misner and H. S. Zapolsky, 1964, Phys. Rev. Lett. {\bf 12}, 635.

\item[]  U. S. Nilsson and C. Uggla, 1997, Class. Quantum Grav. {\bf 14}, 1965.

\item[] A. Ori and T. Piran, 1987, Phys. Rev. Lett. {\bf 59}, 2137.

\item[] A. Ori and T. Piran, 1988, Mon. Not. R. Astron. Soc. {\bf 234}, 821.

\item[] A. Ori and T. Piran, 1990, Phys. Rev. D. {\bf 42}, 1068.

\item[] M. V. Penston, 1969, MNRAS {\bf 144}, 449.

\item[] J. Ponce de Leon, 1988, J. Math. Phys. {\bf 29}, 2479.

\item[] J. Ponce de Leon, 1990, J. Math. Phys. {\bf 31}, 371.

\item[] J. Schwartz, J. P. Ostriker and A. Yahil, 1975, Ap. J. {\bf 202}, 1.
      
\item[] L. I. Sedov, 1967, {\it Similarity and Dimensional Methods in Mechanics} (New York, Academic).

\item[] J. Wainwright and G. F. R. Ellis, 1997, {\it Dynamical Systems in Cosmology} (Cambridge University Press, Cambridge).

\item[] P. Wesson, 1986, Phys. Rev. D {\bf 34}, 3925.

\item[] P. S. Wesson, 1989, Ap. J. {\bf 336}, 58.

\item[] A. Whitworth  and D. Summers, 1985, MNRAS, {\bf 214}, 1.

 \end{enumerate}

\vskip .5in

\noindent
{\large \bf Appendix A.}

In this paper we have used ``comoving'' coordinates, since this
approach is best suited to studying the solutions explicitly.  However, it should be stressed that our work is complemented by the analysis
of Bogoyavlensky (1985) and Goliath et al. (1997, 1998) using
``homothetic'' coordinates and that of Ori \& Piran (1990)
and Maison (1995) using Schwarzschild coordinates. In this Appendix, we discuss these other approaches in more detail.   
 
In the homothetic approach, the coordinates are adapted to the homothetic vector and this
yields results which complement and, in some cases, provide more rigorous  demonstrations of the conclusions reached in this paper. However, in the homothetic
approach, spacetime must be covered by several coordinate patches, one in which
the homothetic vector is spacelike and one in which it is timelike. These
regions must then be joined by a surface in which the homothetic vector is
null and this surface is associated with important physics. 
Bogoyavlensky (1985) studied the spacelike and timelike cases 
simultaneously (with the metric being written in ``conformally static'' form)
and continuously matched the two regions to obtain the behaviour of solutions
crossing the null surface. However, it should be noted that Bogoyavlensky 
changed comoving coordinates explicitly to describe the physics of the associated solutions. 

Recently Goliath et al. (1998a, 1998b) have reinvestigated both the
spatially and temporally self-similar cases. The timelike region  contains the
more interesting physics (eg. shocks and sound-waves). They introduce dimensionless variables, so that the
number of equations in the coupled system of autonomous differential equations
is reduced, with the resulting reduced phase-space being compact and regular.
In this way the similarities with the equations governing hypersurface orthogonal models, and in particular spatially homogeneous models (Wainwright
\& Ellis 1997, Nilsson \& Uggla 1997), can be exploited. In their approach,
all equilibrium points are hyperbolic, in contrast to the earlier work in
which Bogoyavlensky (1985) used non-compact variables (which resulted in parts
of phase-space being ``crushed''). One drawback with this approach is that spacetime must be
covered by more than one patch and the solutions must be matched between the spatially and
temporally self-similar regimes in order to determine their physical properties. 

The Schwarzschild approach is better
suited to studying the
causal structure of the self-similar solutions. This is because, in order to obtain physically
reasonable models, spacetimes are often required to be asymptotically flat. 
Since asymptotically flat spacetimes are not self-similar, one therefore needs to
match a self-similar interior region to an non-self-similar exterior region and
this is usually taken to be Schwarzschild. In particular, Schwarzschild coordinates are most suitable for solving the equations of motion for (radial) null geodesics, as required in studying the global structure of the solution. Consequently it was
used by Ori \& Piran (1990) since one of their primary goals was to study
naked singularities and test the cosmic censorship hypothesis. However, the
Schwarzschild coordinates break down at $t=0$. 

\vskip .5in

\noindent
{\large \bf Appendix B}

The precise transformations 
between the various coordinate systems used to 
study self-similar spherically symmetric perfect fluid 
models are given explictly in Bogoyavlenski (1985; see Section 3 of Chapter IV).  The coordinate 
transformations between the comoving and Schwarzschild 
systems, both of which are employed by Ori \& Piran (1990),
are given explicitly in their paper.  The transformations between 
the homothetic and Schwarzschild coordinates and between the homothetic and comoving coordinates are given explicitly 
in Goliath et al. (1998a; Appendix B), where the relationship 
between their variables and those of Ori \& Piran (1990), 
Maison (1996) and Foglizzo \& 
Henriksen (1992) are also given.
The relationsip between their variables and those used by Bogoyavensky (1985) are given by Goliath et al. (1998).

Here we explicitly demonstrate the relationship
between the variables used in this paper and those used 
in Foglizzo \& Henriksen (1992; FH).  The main
functions used in FH are the three functions ($N$, $\overline{\mu}$, $V^2$),
defined by
eqns (FH3) -- (FH5), which depend on the similarity variable $\xi \equiv z^{-1}$.  The remaining self-similar 
functions can then be written in terms of these
[see eqns (FH6) -- (FH8)].  Their function $V$ is identical to ours. Using eqns (2.12) and (FH3) we find that
$$
N(z) = a_1 x^{(\alpha-1)/\alpha} z^{2(\alpha -1)/(1+\alpha)}. \hfill\qquad (B1)
$$
Using eqns (2.8), (2.12), (2.17) and (FH4), we obtain
$$
\overline{\mu} (z) = 3 \left[1+(1+\alpha) \frac{\dot{S}}{S}\right]. \hfill\qquad (B2)
$$
Conversely, $x$ and $S$ can be defined explicitly in terms of 
$(N, \overline{\mu}, V^2)$ through eqn (FH8),
$$
S^2 = a_2 z^{(\alpha-1)/(1+\alpha)} [N|V|]^{-1},\hfill\qquad (B3)
$$
and eqn (B1).

The differential equations governing the evolution of $(N, \overline{\mu}$, $V^2)$ are given by
eqns (FH12) -- (FH14); these equations constitute an autonomous system
of ODEs in terms of the variable $ln \xi = -ln z$.  Eqns (FH12) and 
(FH13) are equivalent to our eqns (2.15) and (2.16).  The first
integral of the governing ODEs is given 
by eqn (FH10) and is equivalent to our eqns (2.17) and (2.18).
Eqn (FH14), which governs the evolution of $V^2$, 
is obtained by differentiating $V^2$, defined by eqn (2.19), and using 
the first integral. Consequently,
the evolution eqn (FH14) replaces eqns (2.17) and (2.18).

FH then regularize their system of eqns by 
introducing a new independent variable, $\tau$, defined by 
eqn (FH15), which is equivalent to
$$
\frac{d ln z}{d \tau} = -1 + \alpha V^{-2}.  \hfill\qquad (B4)
$$
This divides phase-space into two disconnected components.
Although the resulting system of ODEs is autonomous, 
the system is not regular at $\xi =0$ despite the fact that this point 
does not correspond
to a physical singularity (i.e., it arises due to a coordinate problem at 
$\xi =0$).
FH then introduce new functions and coordinates so that 
solutions are completely regular at $(t =0, r > 0)$. However, the
resulting ODEs are no longer autonomous after this transformation.

\vskip .5in

\noindent
{\large \bf Appendix C}.

The expansion of the comoving fluid velocity congruence, $\theta \equiv u^a \,\!\! _{;a}$,
is given by
$$
r \theta = ze^{-\nu} \Theta \hfill\qquad (C1)
$$
where
$$
\Theta (z) \equiv   - \frac{d}{dz} (\lambda + 2S). \hfill\qquad (C2)
$$
When $p = \alpha \mu$, the conservation equations then yield
$$
\frac{dW}{dz} = -(1+\alpha) W\Theta. \hfill\qquad (C3)
$$
If we consider two comoving perfect fluids as the source of the gravitational 
field, each of which satisfies
$$
p_i = \alpha_i \mu_i, \;\; W_i = \mu_i R^2 \;\;( \alpha = 1, 2) \hfill\qquad (C4)
$$
(e.g., a mixture of dust and radiation with $\alpha_1 = 1/3$ and $\alpha_2 =0$),
then the source is equivalent to a single perfect fluid with
$$
\mu = \mu_1 + \mu_2,\;\;\;
p = p_1 + p_2 = \alpha_1 \mu_1 + \alpha_2 \mu_2 \hfill\qquad (C5)
$$
although this  does not admit an equation of state.

Suppose the two perfect fluids are non-interacting, with each separately 
satisfying the conservation equation (C3). Then
$$
\frac{dW_i}{dz} = -(1+ \alpha_i) W_i \Theta. \hfill\qquad (C6)
$$
We define a new variable, $\chi$, by
$$
\chi = \chi (z) \equiv  \frac{\mu_1 - \mu_2}{\mu_1 + \mu_2} = \frac{W_1 - W_2}{W_1 + W_2}, \hfill\qquad (C7)
$$
where $-1 \leq \chi \leq 1$.  From eqn (C6) we derive
the evolution equation for $\chi$:
$$
\frac{d \chi}{d \tau} = \frac{1}{2} (\alpha_1 - \alpha_2) \{1 - \chi^2\} \hfill\qquad (C8)
$$
where $\tau$ is defined by
$$
\frac{d \tau}{dz} = - \Theta  \hfill\qquad (C9)
$$
for regions in which $\Theta$ (and hence the 
expansion $\theta$) is non-zero.

Eqn (C8) is a decoupled autonomous equation for $\chi$. It has equilibrium points $\chi = \pm 1$, and 
hence all solutions asymptote to $\chi = \pm 1$ in 
regions for which the expansion does not become zero.  
$\chi = + 1$ corresponds 
to $\mu_2 =0$ and $\chi =-1$ corresponds to $\mu_1 = 0$; i.e., the 
solutions of these self-similar two-fluid models asymptote
towards the exact asymptotes of the single perfect fluid solutions. Asymptotically
$$
p/\mu \rightarrow \alpha_i, \hfill\qquad (C10)
$$
and which value of $\alpha_i$ is picked out (i.e., which of the two 
single fluids govern the dynamics asymptotically) depends on the signs
of $(\alpha_1 - \alpha_2)$ and $\Theta$ and on whether $|z| \rightarrow 0$ or $|z| \rightarrow \infty$. 
 
\vskip .5in

\noindent
{\large \bf Figures}

FIGURE (1). This shows the form of $V(z)$ for the exact Friedmann (F), static (S) and Kantowski-Sachs (KS) solutions for (a) the general $\alpha$ case and (b) the $\alpha=1/3$ case. Also shown are the sonic lines $|V|=1/\sqrt{\alpha}$ (broken), the range of values of $z$ in which curves can be regular at the sonic point ($z_1<z<z_2$ and $z>z_3$) and (for $\alpha =1/3$) the curves corresponding
to $M=0$ (solid).
 
FIGURE (2). This shows the form of the scale factor $S(z)$, the velocity function $V(z)$ and the density function $\mu t^2(z)$ for the asymptotically Friedmann dust models. These are described by 
a single parameter $E$ where $E=0$ in the exact Friedmann case: the $z>0$ solutions are overdense and collapse to black holes for $E<0$ (with an event horizon for $E>E_*$ since $V_{min}<1$) but they are underdense and expand forever for $E>0$. The $z<0$ solutions 
are just the time reverse of these. The broken curve corresponds to a solution with negative mass and is probably unphysical. 

FIGURE (3). This shows the form of the scale factor $S(z)$, the velocity function
$V(z)$ and the density function $\mu t^2(z)$ for dust models with $E=0$ and different values of $D$. These solutions necessarily span both positive and negative values of $z$. For $D>0$ they represent expanding (solid) or collapsing (broken) solutions and the latter contain a naked
singularity ($V_{min}<1$) if $D$ exceeds $D_*$. The $D<0$ models (dotted) undergo shell-crossing before encountering the singularity and are probably unphysical.

FIGURE (4). This shows the form of the scale factor $S(z)$, the velocity function
$V(z)$ and the density function $\mu t^2(z)$ for the asymptotically quasi-static
dust solutions. These are described by two parameters ($D$ and $E$). For $E>0$ they
resemble the curves in Figure (3), with both expanding (solid) and collapsing (broken) solutions. The collapse singularity is naked ($V_{min}<1$) if 
$E$ exceeds a value $E_*(D)$. For $E<0$ there are also solutions which recollapse to black holes or emerge from white holes (broken), as in the asymptotically Friedmann case. The
dotted curve corresponds to an (unphysical) negative mass solution.

FIGURE (5). This shows the form of the asymptotically Friedmann solutions for a radiation equation of state ($\alpha=1/3$),
with particular emphasis on the behaviour at the sonic point. Solutions which are 
regular (irregular) at the sonic point are shown by solid (dotted) lines, while black hole solutions (with no sonic point) are shown by broken lines. There is a
1-parameter continuum of underdense solutions which are regular at the sonic point but the overdense solutions lie in discrete bands (just the first of which is shown) and are characterized
by the number of oscillations they exhibit. 

FIGURE (6). This shows the form of the velocity function $V(z)$ for the
full family of spherically symmetric similarity solutions with $\alpha =1/3$.  The exact Friedmann, Kantowski-Sachs and static solutions are indicated by the bold lines. Also shown are the asymptotically Friedmann solutions (for different values of $E$) and the asymptotically quasi-static solutions (for different values of $E$ and $D$). The asymptotically quasi-static solutions contain a naked singularity when the minimum of $V$ is below 1. The negative  $V$ region is occupied by the asymptotically Kankowski-Sachs solutions, though these may not be physical since the mass is negative; solutions which are irregular at the sonic point are shown by broken lines. Also shown are some of the solutions which are  asymptotically Minkowski at finite and infinite values of $z$.

FIGURE (7). This shows the form of the velocity function $V(z)$ for the
full family of spherically symmetric similarity solutions with $\alpha < 1/3$.

FIGURE (8). This shows the form of the velocity function $V(z)$ for the
full family of spherically symmetric similarity solutions with $\alpha > 1/3$.
\vskip .5in

\end{document}